\documentclass[graybox,vecphys]{svmult}

\usepackage[]{amsmath}

\usepackage{mathptmx}       
\usepackage{helvet}         
\usepackage{courier}        
\usepackage{type1cm}        
\usepackage{makeidx}         
\usepackage{graphicx}        
\usepackage{multicol}        
\usepackage[bottom]{footmisc}

\makeindex             

\begin{document}

\title*{Unstable Periodically Forced Navier-Stokes Solutions---Towards Nonlinear First-Principle  Reduced-Order Modeling of Actuator Performance}

\titlerunning{Towards Nonlinear First-Principle Reduced-Order Modeling of Actuator Performance} 
\authorrunning{⟨Marek Morzy\'nski et al.⟩}

\author{Marek Morzy\'nski \and Wojciech Szeliga \and Bernd R.~Noack}


\institute{Marek Morzy\'nski \at Pozna\'n University of Technology, Chair of Virtual Engineering, Jana Paw\l{}a II 24, PL 60-965 Pozna\'n, Poland, \email{morzynski@virtual.edu.pl}
\and Wojciech Szeliga \at Pozna\'n University of Technology, Chair of Virtual Engineering, Jana Paw\l{}a II 24, PL 60-965 Pozna\'n, Poland; 
Pozna\'n Supercomputing and Networking Center, Jana Paw\l{}a II 10, PL 61-139 Pozna\'n, Poland,
\email{wojciech.s.szeliga@doctorate.put.poznan.pl}
\and Bernd R.~Noack \at LIMSI-CNRS, UPR 3251, 
Rue John von Neumann, Campus Univeristaire d'Orsay, B\^at 508, F91405 Orsay, France; 
Institute for Turbulence-Noise-Vibration Interactions and Control,
Harbin Institute of Technology,
Graduate School Shenzhen, HIT Campus, University Town, Xili, 58800 Shenzhen, China;
Hermann-F\"ottinger Institute, Technische Universt\"at Berlin, Stra{\ss}e des 17.\ Juni 135, D-10623 Berlin, Germany
\email{Bernd.Noack@limsi.fr}}
%
%
\maketitle

\abstract*{We advance the computation of physical modal expansions for unsteady incompressible flows.
Point of departure is a linearization of the Navier-Stokes equations around its fixed point
in a frequency domain formulation.
While the most amplified stability eigenmode is readily identified by a power method,
the technical challenge is the computation of more damped higher-order eigenmodes.
This challenge is addressed by a novel method to compute unstable periodically forced solutions
of the linearized Navier-Stokes solution.
This method utilizes two key enablers.
First,  the linear dynamics is transformed by a complex shift of the eigenvalues 
amplifying the flow response at the given frequency of interest.
Second, the growth rate is obtained from an iteration procedure.
The method is demonstrated for several wake flows 
around a circular cylinder,
a fluidic pinball, i.e.\ the wake behind a cluster of cylinders,
a wall-mounted cylinder,
a sphere and a delta wing.
The example of flow control with periodic wake actuation and forced physical modes paves the way for applications of physical modal expansions.
These results encourage Galerkin models of three-dimensional flows utilizing Navier-Stokes based modes.
} 

\abstract{We advance the computation of physical modal expansions for unsteady incompressible flows.
Point of departure is a linearization of the Navier-Stokes equations around its fixed point
in a frequency domain formulation.
While the most amplified stability eigenmode is readily identified by a power method,
the technical challenge is the computation of more damped higher-order eigenmodes.
This challenge is addressed by a novel method to compute unstable periodically forced solutions
of the linearized Navier-Stokes solution.
This method utilizes two key enablers.
First,  the linear dynamics is transformed by a complex shift of the eigenvalues 
amplifying the flow response at the given frequency of interest.
Second, the growth rate is obtained from an iteration procedure.
The method is demonstrated for several wake flows 
around a circular cylinder,
a fluidic pinball, i.e.\ the wake behind a cluster of cylinders,
a wall-mounted cylinder,
a sphere and a delta wing.
The example of flow control with periodic wake actuation and forced physical modes paves the way for applications of physical modal expansions.
These results encourage Galerkin models of three-dimensional flows utilizing Navier-Stokes based modes.} 


\section{Introduction}
\label{sec:1}

Modeling instabilities and subsequent soft or hard bifurcations
is at the heart of understanding laminar, transitional and turbulent flow dynamics.
Many prominent coherent structures arise from instabilities.
Well studied examples are the von K\'arm\'an vortex street behind cylinders,
the Kelvin-Helmholtz vortices of mixing layers,
the Tollmien-Schlichting waves in laminar boundary layers,
the Taylor-Couette vortices between two centered rotating cylinders,
the G\"ortler vortices in a boundary layer of an inward curved surface,
and the Raleigh-B\'enard cells in a horizontal plane fluid layer heated from below,
just to name a few. 

Many of these coherent structures also characterize the turbulent flow
at Reynolds numbers which are orders of magnitudes beyond the bifurcation values.
These observations have fueled a hypothesis in the 1940's that turbulence
may be the infinite succession of Hopf bifurcations  \cite{Landau1944,Hopf1948}.
In the 1980's other transition scenarios have been added leading to strange attractors.
Prominent examples are the Feigenbaum \cite{Feigenbaum1978jsp}, 
Ruelle-Takens \cite{Ruelle1971cmp}, 
Ruelle-Takens-Newhouse \cite{Newhouse1978cmp} and 
Manneville-Pomeau scenario \cite{Pomeaou1980cmp}.
The list of transition scenarios could easily be extended.

The engineering relevance of these pioneering works to turbulence is rather limited,
because the applicability is restricted to few configurations and a narrow Reynolds number regimes.
Yet, these studies have provided important qualitative insights into the flow dynamics,
like the notion of divergent trajectories, strange attractors, manifolds, 
domains of attraction, edge states, just to a name a few.
In addition, the instabilities of laminar flows, 
like boundary-layer transition, 
is quantitatively characterized by stability analysis
and has important engineering applications.

The interest in modal analysis of flows is twofold. 
Modal analysis is a tool for investigation of flow phenomena. 
On the other hand it is of vital interest for flow control.  
Physical modes were characterized above---there are modes which we are able to trigger in the flow field with an actuation. 
If the actuation is random and small 
the flow responds with the leading eigenmodes.
However, the flow may respond to strong actuation 
with non-leading and sometimes local eigenmodes.  
Control strategies often take advantage 
of non-leading modes to control the dominating ones via frequency crosstalk. 
In this contribution, 
such a technique will be encapsulated in a nonlinear first-principle reduced-order model
using tools of stability analysis. 
The use of physical modes---the leading and the non-leading ones 
is even more important for closed-loop, model-based control. 
To build a controller, the high- or infinite-dimensional system is projected onto a set of modes --- of empirical, physical or mathematical type --- to obtain a least-dimensional representation. 
The use of physical modes in this situation may be more effective 
to represent flow response 
than employing a large number of mathematical or empirical modes. 

Another enabling aspect of stability analyses 
is that the resulting eigenmodes may be suited for Galerkin models
describing flow dynamics beyond the assumptions of bifurcation theory.
One example is the Galerkin model 
over a separated boundary layer \cite{Akervik2007jfm}
or vortex shedding behind a wake \cite{Noack2003jfm}.
For simple configurations,
stability eigenmodes have be shown 
to form a basis for arbitrary flows \cite{Grosch1978jfm,Salwen1981jfm}
or distill at least the dominant coherent structures.
Alternative data-driven to first-principle-based expansions are discussed in  \cite{Taira2018arXiv}.
The dominant one-global stability eigenmodes are easy to compute
while Bi- and Tri--global modes---following the terminology of Theofilis \cite{theofilis2003advances,Theofilis2011arfm} (meaning fully two--dimensional or fully three--dimensional) stability analysis provide a significant challenge.

This study is devoted to advancing 
the computation on several stability eigenmodes
for complex three-dimensional flows.
Our investigation of linearized Navier-Stokes equations 
assesses either global modes of the flow or responses to a particular perturbation. 
Contrary to receptivity studies, we analyze Navier-Stokes perturbation equation in frequency domain 
and apply a complex shift to narrow down the investigated frequency range. 
The analysis employs elements of our earlier methods of global (fully two--dimensional or fully three--dimensional) flow stability investigations, elements of our shift- inverse \cite{ZAMM89}, subspace methodology \cite{MORZYNSKI1999} and three-dimensional numerical solutions \cite{Morzynski2008gdansk}. 
It is following the spirit and directions of \cite{gomez2015use,LIU2016316,semeraro_bagheri_brandt_henningson_2011, 
garnaud_lesshafft_schmid_huerre_2013,åkervik_hœpffner_ehrenstein_henningson_2007,chomaz2005global}.

The manuscript is outlined as follows:
Section \ref{sec:2} details the proposed methodology
starting from frequency domain formulation
of the linearized Navier-Stokes equations.
Sections \ref{sec:3} and \ref{sec:4} describe investigations of two- and three-dimensional wake flows, respectively.
Section \ref{sec:5} shows the example of reduced-order modeling of periodic wake actuation with the technique of forced physical modes, developed in previous sections.
In the conclusions (Sec.~\ref{sec:6}), we summarize the study and outline perspectives for future work.

\section{Frequency Domain Formulation and Solution}
\label{sec:2}

\subsection{Navier-Stokes Equations in Frequency Domain}
\label{subsec:21}

We focus on incompressible flows around obstacles in a steady domain $\Omega$.
The Newtonian fluid is characterized by the constant density $\rho$ and kinematic viscosity $\nu$.
The flow is described in a Cartesian coordinate system ${\bf x}=(x,y,z)=(x_1,x_2,x_3)$
with corresponding velocity vector ${\bf V} = (V_1,V_2,V_3)$ and pressure $P$.
The time is represented by $t$ and the volume force by ${\bf F} = (F_1,F_2,F_3)$.
In the sequel, all quantities are assumed to nondimensionalized
with density $\rho$, the characteristic velocity $U$ and characteristic length $L$.
Thus, the flow solution is parameterized by the Reynolds number 
\begin{equation}
\label{Eqn:Re}
Re=\frac{UL}{\nu}.
\end{equation}

The evolution is  described by the Navier-Stokes equations 
\begin{equation}
\dot{V}_{i}+V_{i,j}V_{j}+P_{,i}-\frac{1}{Re}V_{i,jj}= F_i 
\label{Eqn:NS}
\end{equation}
and equation of continuity
\begin{equation}
{V}_{i,i}=0.
\end{equation}
Here the dot denotes temporal differentiation,
the superscript `$,i$' denotes partial differentiation with respect to $x_i$
and `$,ii$' double partial differentiation with respect to $x_i$.

The steady solution of \eqref{Eqn:NS}
is indicated by a bar, i.e.\
$\bar{V}_i$ and $\bar{P}$ denote the velocity component
and pressure under the steady force $\bar{F}$.
We decompose the flow variables in a steady component
and disturbance indicated by an acute symbol: 
\begin{subequations}
\begin{eqnarray}
V_{i} & = & \bar{V_{i}}+\acute{V_{i}}\\
P & = & \bar{P}+\acute{P}  \\
F & = & \bar{F}+\acute{F}.
\end{eqnarray}
\label{Eqn:Decomposition}
\end{subequations}

The evolution equation of the disturbance 
is derived from substituting \eqref{Eqn:Decomposition}
in \eqref{Eqn:NS} and subtracting the steady Navier-Stokes equations.
The corresponding Navier-Stokes equation and equation of continuity read
\begin{equation}
\dot{\acute{V}_{i}}+\acute{V}_{j}\bar{V}_{i,j}+\bar{V}_{j}\acute{V}_{i,j}+\acute{V}_{j}\acute{V}_{i,j}+\acute{P}_{,i}-\frac{1}{Re}\acute{V}_{i,jj}=\acute{F}_i 
\label{Eqn:Dist}
\end{equation}
and
\begin{equation}
\acute{V}_{i,i}=0.
\end{equation}

Equation \eqref{Eqn:Dist} has linear terms in velocity, pressure and force
and one quadratic term in the velocity.
Close to the steady solution, the quadratic term $\acute{V}_{j}\acute{V}_{i,j}$ can be ignored,
yielding the linearized Navier-Stokes equations.

In the next step, the disturbance is subjected to the 
separation ansatz in time and space dependence: 
\begin{eqnarray}
\acute{V}_{i}(x,y,z,t) & = & \tilde{V}_{i}(x,y,z)\,e^{\lambda t} +c.c.\label{expon}\\
\acute{P}(x,y,z,t) & = & \tilde{P}(x,y,z)\,e^{\lambda t} +c.c. \nonumber \\
\acute{F}_{i}(x,y,z,t) & = & \tilde{F}_{i}(x,y,z)\,e^{\lambda t} +c.c. \nonumber 
\end{eqnarray}
where $e^{\lambda t}$ represents the temporal dynamics with complex eigenvalue $\lambda$
and the tilde marks the spatial amplitude of the mode.
Now, the linearized evolution equations in the frequency
domain read:
\begin{eqnarray}
\lambda\tilde{V}_{i}+\tilde{V}_{j}\bar{V}_{i,j}+\bar{V}_{j}\tilde{V}_{i,j}+\tilde{P}_{,i}-\frac{1}{Re}\tilde{V}_{i,jj}=\tilde{F}_i \nonumber \\
\tilde{V}_{i,i}=0 . \label{eq:diffEV}
\end{eqnarray}
The homogeneous form of equation \eqref{eq:diffEV} with vanishing force 
constitutes the generalized complex, nonhermitian eigenvalue problem. 
We use the penalty formulation of the Navier-Stokes equations
for the examples presented here, i.e.\ for the steady base solution, eigensolutions, and unsteady computations. 
The penalty method eliminates the pressure from governing equations and enforces  near incompressibility via:
\begin{equation} \label{penalty}
\frac {1} {\epsilon} P = V_{j,j}
\end{equation}
The
penalty parameter  $\epsilon$ is usually chosen to be a large number, related
to computer word length.   In the calculations
presented in the following sections we assume $\epsilon = 10^5$ and keep it
constant in the iteration process. The Ladyzenskaja-Babu{\v{s}}ka-Brezzi (LBB)
condition stating the
allowable interpolating functions for velocity and pressure is, in the penalty
formulation, equivalent to using a reduced integration of the penalty term.
It should be mentioned that in other studies, including flow computations and eigenanalysis, we have used also the velocity-pressure Finite Element Method (FEM) formulation of the Navier-Stokes equations, obtaining almost identical results. 

In compact form the equation \eqref{eq:diffEV} can be written as
\begin{equation}
\bf{Aq} + \lambda  \bf{Bq} = F,
\label{VEWP-1-2-1}
\end{equation}
where $\bf{q}$ represents the flow variables $\tilde{V}$ and $\tilde{P}$,
and $\bf{F}$ comprises $\tilde{F}_i$. 
From the considered solutions of equations \eqref{eq:diffEV} to \eqref{VEWP-1-2-1} 
we exclude trivial ones.
For all practical purposes, 
$\bf{q}$ and $\bf{F}$ denote the flow and force discretization
on a grid, $\bf{A}$ represents the matrix associated with the linear Navier-Stokes dynamics
and $\bf{B}$ the mass matrix.

After a realistic discretization of equations
\eqref{eq:diffEV},
 the number of degrees of freedom (DOF), i.e.\ the dimension of $\bf{q}$, 
is usually very large, 
at least of order of $O\left(10^{5}\right)$ for two-dimensional 
and $0\left( 10^{6}\right)$ for the three-dimensional problem.
The corresponding eigenvalue problem constitutes a numerical challenge. 
Many eigenvalues $\lambda$ form  complex conjugate pairs 
which causes additional
problems for solution algorithms.

Instead of direct solution of the homogeneous eigenvalue problem,
we investigate the flow response in frequency domain. 
In contrary to commonly employed ''black box''approaches 
we shall take advantage of knowledge of governing equations. 

It is practical to split the complex equation \eqref{eq:diffEV} into
two real-valued equations which read:
\begin{eqnarray}
{\bf Aq}_{Re} + \lambda_{Re} {\bf Bq}_{Re} + \lambda_{Im} {\bf Bq}_{Im}&=&{\bf F}_{Re}\nonumber \\
{\bf Aq}_{Im} + \lambda_{Re} {\bf Bq}_{Im} - \lambda_{Im} {\bf Bq}_{Re}&=&{\bf F}_{Im}. \label{eq:EV_Re_Im}
\end{eqnarray}
The inverse of the complex system matrix of \eqref{eq:EV_Re_Im} plays
the role of frequency response function relating the amplitudes and
the phases of the system response with the forcing function.

\subsection{Boundary Conditions}

For the steady base solution, on the inflow,
and on the upper, lower, front and back boundary, we set the Dirichlet boundary conditions:
\begin{equation}
 V_x = 1, \; V_y=0, \; V_z=0
\end{equation}
where $x$ is the direction of the inflow.
On
the outflow of the domain we use the condition:
\begin{equation}
\int_\Gamma \Phi_m \sigma_{ij} n_j \, d\Gamma = 0
\end{equation}
where $\Phi$ is FEM weighting function, $\sigma_{ij}$ denotes the stress tensor and $n_j$ is the unit vector normal to the boundary $\Gamma$. It is the ''do nothing'', homogeneous Neumann,   
natural, non-reflecting boundary condition for the Finite Element Method, fully ``transparent'' to the flow.
On the body, the  no-slip boundary condition reads:
\begin{equation}
 V_x = 0, \; V_y=0, \; V_z=0.
\end{equation}

For eigensolutions the homogeneous Dirichlet boundary conditions:
\begin{equation}
 \tilde{V_x}= 0, \; \tilde{V_y}=0, \; \tilde{V_z}=0
\end{equation}
are set on the inflow, upper, lower, front and back boundaries. On the outflow the natural
boundary conditions, analogous with those applied for steady flow, are
introduced.
The forced actuation in the frequency domain is equivalent to either Dirichlet boundary condition or inhomogeneous Neumann boundary condition.
For example, the oscillatory motion of the cylinder, having diameter $D=1$, is introduced via setting
\begin{eqnarray}
 V_x &=&   \quad 2 A \, y_C
\\
\nonumber
  V_y &=&  \, -2 A \, x_C
\end{eqnarray}
where $A$ is the amplitude of the oscillation and $x_C$ and $y_C$ are the coordinates of a node on the cylinder surface.
Inhomogeneous volume force can be set in the domain as the non-zero entrance to right hand side vector (RHS) of the Finite Element Method system.  
The indices and boundary conditions for the two-dimensional examples are limited only to the $x,y$ values.

\subsection{Complex Shift}
\label{subsec:22}

\begin{figure}
\begin{centering}
\includegraphics[scale=0.2]{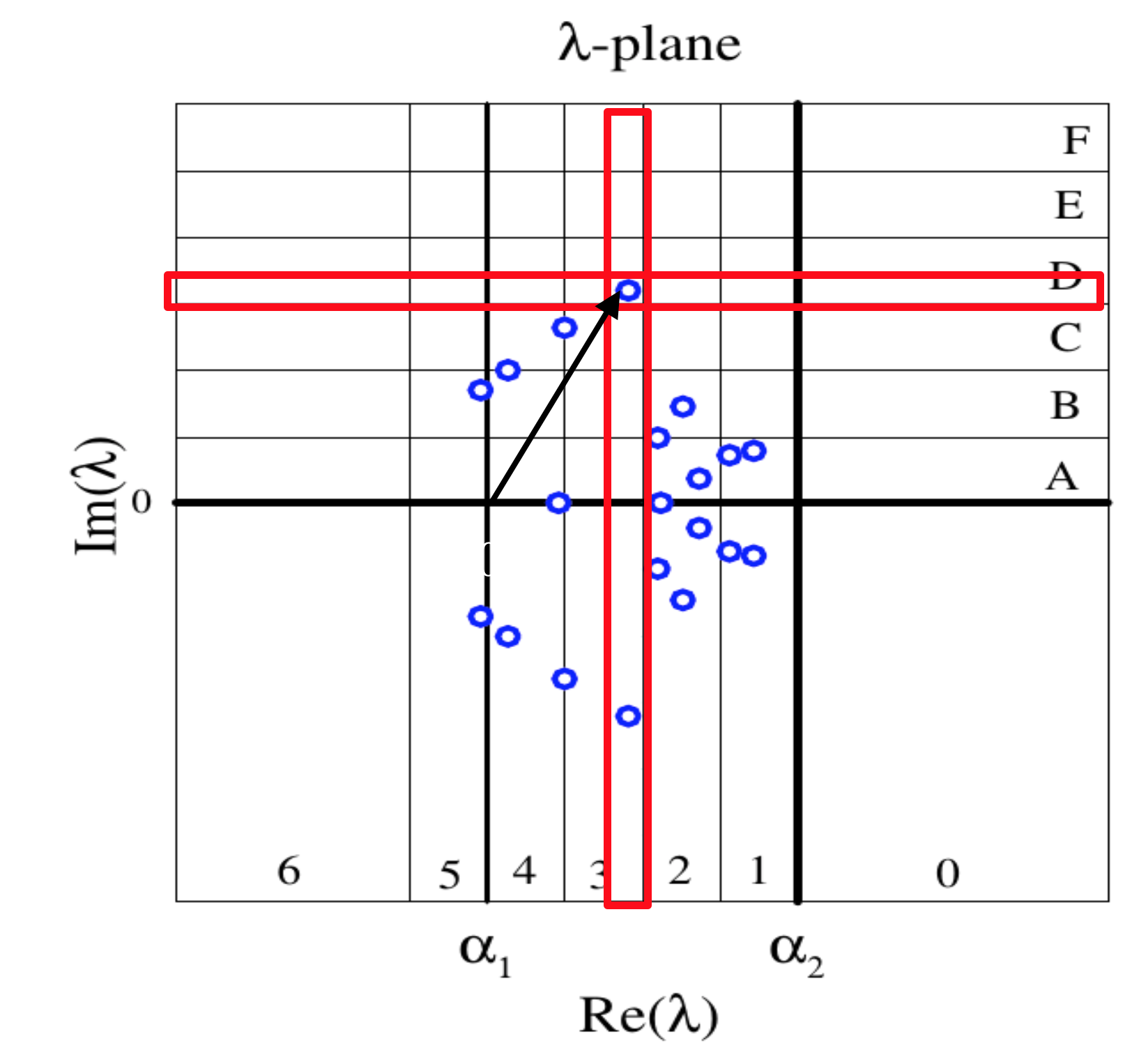}
\par
\end{centering}
\caption{Complex shift for the eigenvalue spectrum of a circular cylinder at $Re=100$, adapted from  \cite{MORZYNSKI1999}.}
\label{fig:Complex_shift}
\end{figure}
Let us assume that the approximate eigenvalue of interest $\mu$ is known 
and $\alpha$ is the difference to the accurate eigenvalue $\lambda$
\begin{equation}
\label{Eqn:LambdaShift}
\lambda=\mu+\alpha 
.
\end{equation}
In this case,  original eigenvalue problem (\ref{VEWP-1-2-1})
may be shifted 
\begin{equation}
(\bf{A} + \alpha \bf{B}){\bf q} + \mu {\bf Bq} = {\bf 0 }  ,
\label{VEWP-1-2-1-1}
\end{equation}
where we solve for the small error $\alpha$.
Equation \eqref{Eqn:LambdaShift} allows us to reconstruct
the eigenvalues for original system (\ref{VEWP-1-2-1}).
The eigenvectors of original and shifted eigensystem are the same.

There are many advantages in the use of the shifted eigenvalue problem. 
The convergence of the approximated eigenvalue problem is much faster 
as it is for the original eigenvalue of the system.
Second, the singularity or ill conditioning of the system matrix is avoided.
Third and most importantly, we can choose the ''area of interest'' in the eigenvalue spectrum.

This methodology is demonstrated in Fig.~\ref{fig:Complex_shift} 
and details were presented in \cite{MORZYNSKI1999}. 
The figure shows the eigenvalues for the circular cylinder flow at $Re=100$ in the complex plane. 
The horizontal, real axis refers to the growth rate of the eigenvalue  
while the vertical one corresponds to it's frequency. 
With the complex shift, indicated in the figure by the arrow we can compute the eigensolution for arbitrary eigenvalues.
This allows us to skip the most amplified eigenvalues, 
related to the von K\'arm\'an mode and focus  on the other.

The technique of the complex shift can be also used if 
want to investigate the frequency transfer function for particular
frequency and growth rate of the disturbance and this approximate
is equal to $\alpha$. 
\begin{equation}
(\bf{A} + \alpha \bf{B}){\bf q} + \mu {\bf B q} = {\bf F }.
\label{VEWP-1-1-2}
\end{equation}

The equation (\ref{VEWP-1-1-2}) represents the fluid flow in which all the oscillations except the ones close to $\alpha$ are 
eliminated or damped. 
The dominant mode (e.g. von K\'arm\'an mode for bluff bodies) 
which usually dominates and overrides all other modes is not triggered 
and we can find the (linear) response of shear layer or boundary layer 
to the actuation even if the corresponding eigenvalues are damped.
This information can be relevant for the flow control 
as we can estimate the growth rate of the mode 
and the energy which should be delivered to the actuator for this mode to pop up. 
Actuating the additional modes can be  essential in damping of the dominating, most energetic one. 
We can also optimize the approximate (linear) flow response as a function of actuator's placement.

\subsection{Structural Mechanics Analogy}
\label{subsec:23}

Structural mechanics was one of the first technical fields 
which used modal analysis and reduced-order models (ROM). 
It employs methods and tools like static and dynamic condensation, 
modal coordinates reduction, 
component mode synthesis and the Ritz vector method. 
The latter is used to generate modes, approximate eigenvectors or define the transfer function of the system. 
All these problems are also important in fluid dynamics.

In the structural mechanics formulation, 
the homogeneous form of equation \eqref{eq:EV_Re_Im}
is usually written as: 
\begin{equation}
{\bf K x}  - \omega_{c}^{2} \, {\bf M  x } = {\bf 0} ,
\end{equation}
where ${\omega_{c}^{2}}$ is equivalent to $-\lambda$, 
$\bf K$ denotes stiffness matrix  and $\bf M$  is the mass matrix.

The solution method can be described as Ritz vector method.
It has been used mainly in structural mechanics as the alternative
or complement of modal approach \cite{qu2013model}. In structural mechanics
we apply static Ritz Vector Methods (RVM),
often referred to as Wilson-Yuan-Dickens (WYD) approach \cite{wilson1982dynamic}. 
WYD employs sequential solution of:
\begin{equation}
\hat{{\bf x}}_{1}={\bf K}^{-1}{\bf g}
.
\end{equation}
The inertia term, neglected in the first step is used to generate the new Ritz vectors: 
\begin{equation}
\hat{{\bf x}}_{i}={\bf K}^{-1}{\bf M}\hat{{\bf \, x}}_{i-1}
\end{equation}
with subsequent $\bf M$-normalization
\begin{equation}
{\bf x}_{1}=\frac{\hat{{\bf x}}_{1}}{(\hat{{\bf x}}_{1}^{T}{\bf M}\,\hat{{\bf x}}_{1})^{\frac{1}{2}}}
\end{equation}
and $\bf M$-orthonormalization
\begin{equation}
{\bf \hat{{\bf x}}}_{i}=\tilde{{\bf x}}_{i}-\sum_{j=1}^{i-1}({\bf x}_{j}^{T}{\bf M}\tilde{{\bf x}}_{i}){\bf x}_{i-1}.
\label{eq:Zu}
\end{equation}

Lanczos procedure is also one of the possible choices. 
WYD looses its performance for larger amount of not orthogonal vectors. 
To alleviate this difficulty, the Load-Dependent-Ritz Vectors method (LDRV) is extended to so called quasi-static Ritz
vector methods in which Ritz vector span the model space at desired
prescribed frequencies to efficiently capture the dynamic response
of the system at desired frequency \cite{qu2013model} :
\begin{eqnarray}
\hat{{\bf x}}_{1}&=&({\bf K} - \omega_{c}^{2} \, {\bf M})^{-1}{\bf g} \nonumber
\\
{\bf \tilde{{\bf x}}}_{i}&=&({\bf K} - \omega_{c}^{2} \,{\bf M})^{-1}{\bf M}{\bf x}_{i-1}.
\label{eq:Zu-volume_force}
\end{eqnarray}

The first step obtained with this procedure represents the frequency response to the initial load. 
In the second step,
the frequency response ${\bf x}_{2}$ of the system is enriched by additional inertial
force of ${\bf M}{\bf x}_{i-1}$. 
After mass orthonormalization
the new Ritz vector set spans a wider space of the dynamic response
of the system \cite{qu2013model}.
The algorithms presented here can be used for finding the eigenvectors of the system.

There are similarities in the solution and transformation used in the 
presented procedure and iterative eigenvalue solvers. Similar procedures are applied for inverse iteration, subspace iteration and for the eigenvalue shift. 
For example, the subspace iteration forces the system with initial set of orthogonal 
vectors, which are becoming Ritz vectors during the iteration and eigenvectors at the end of the procedure. For generalized 
eigenvalue problem the $B$ matrix is used to transform the Ritz vectors to a 
new set, used in the next iteration. We use a single vector and do not compute the 
eigenvalue with the reduced model build on the space spanned by the Ritz vectors. 
The algorithm is very similar to Arnoldi iteration but uses shifted
and inverted matrix and is applied to a generalized problem. 

\subsection{The Solution Algorithm}
\label{subsec:24}
The above described methodology can be directly applied to create a modal basis 
and ROM of the fluid flow. 
Here, we adopt ideas of Quasi-Static Ritz Vector Method 
and Load-Dependent-Ritz Vectors method in the frequency domain. 
The methods were briefly introduced in Sect.~\ref{subsec:23}. LDRV method was first introduced by Wilson Yuans and Dickens in \cite{wilson1982dynamic} in structural dynamics. Ritz Vector Methods are generally used as the alternative to mode superposition. In this method the Ritz vectors are generated by loads, which translate in fluid dynamics to volume forces. Generation of load dependent Ritz vectors is far less  computationally expensive than finding the eigenvectors and still spans the space adequate for accurate models. 
 
In our approach the actuation of the flow is implemented with volume forces, 
acting either in discrete points or ones randomly distributed in the flow volume. 
Actuation can be also introduced as the Dirichlet boundary condition. 
It can be applied only in the first iteration, 
to trigger the flow response or can be held constant in all the iterations.
Our system is complex and is characterized by frequency and growth rate. 
This distinguishes the approach from Static Ritz Vector method described above. 
We are forcing fluid motion not only at given frequency 
as in case described by equation \eqref{eq:Zu-volume_force}
but also prescribe the real part of the eigenvalue damping or amplifying
the fluid response. 

In terms of eigenvalue solvers, we apply here the
complex shift to the eigensystem of equations.

We disturb the flow with the volume force ${\bf Bq}_{Re}^{F}$ and ${\bf B q}_{Im}^{F}$.
Here, $\bf B$ can be considered the gain of a small random velocity  ${\bf q}^{F}$  
with support in the whole computational domain or a single point forcing. 
It should be noted that the introduced random disturbance has the frequency determined by $\lambda_{Im}$.
\begin{eqnarray}
{\bf Aq}_{Re}+\lambda_{Re} {\bf Bq}_{Re}+\lambda_{Im}{\bf Bq}_{Im}&=&-{\bf Bq}_{Re}^{F}\nonumber \\
{\bf Aq}_{Im}+\lambda_{Re} {\bf Bq}_{Im}-\lambda_{Im}{\bf Bq}_{Re}&=&-{\bf Bq}_{Im}^{F} 
.
\end{eqnarray}
The iteration with respect to $i$ is performed with
\begin{equation}
{\bf A q}^{(i)} + \lambda {\bf B q}^{(i)} = - {\bf B q}^{(i-1)} \label{eq:EV_Re_Im-1}.
\end{equation}
In the equation \eqref{eq:EV_Re_Im-1} we skip, for brevity of notation, the decomposition of complex equation to the two real number ones. The equation \eqref{eq:EV_Re_Im-1} implies the solution procedure. Once the initial volume force distribution is given it is placed in the right-hand side (RHS) of the equation \eqref{eq:EV_Re_Im-1}. Also initial velocities, if given, can be transferred to the force value via the matrix $\bf B$ and placed in the RHS vector. The next step is the solution of the linear system of equations, giving the flow field response to the applied actuation. There is however another factor which has to be taken into account. For $\lambda$ we assumed the imaginary part $\lambda_{Im}$, being the angular frequency of the actuation. The real part, $\lambda_{Re}$ is unknown, being the parameter which has to be iteratively adjusted by the procedure. For this reason we must use an iterative approach, repeating the solution several times. To solve this problem we adopted the Newton-Raphson method implemented in our nonlinear Navier-Stokes solver:
\begin{eqnarray}
{\bf R } \left( {\bf q}^{(i-1)}+\Delta {\bf q} \right) &=& {\bf R}( {\bf q}^{(i-1)})+{\bf J}^{(i-1)}\Delta {\bf q} 
\\
{\bf q}^{(i)} &=& {\bf q} ^{(i-1)} - [ {\bf J}^{(i-1)}]^{-1} {\bf R} ({\bf q}^{(i-1)}),
\end{eqnarray}
where $\bf R$ represents the residuum of the equation \eqref{eq:EV_Re_Im-1}, 
 dependent on 
iterated $\lambda_{Re}$
and $\bf J$ is the Jacobian of the left-hand side of \eqref{eq:EV_Re_Im-1}.
Again, we replace the complex computations with the real arithmetic.

The real shift $\lambda_{Re}$ is adjusted by the solver to preserve 
the amplitude of $\bf q$ in subsequent iterations.
This task usually takes few iterations.
A good guess of $\lambda_{Re}$ is not necessary
but significantly speeds up computations. 
As the convergence is obtained the artificial zero-growth rate shift
is equal to $1+\lambda_{Re}$.
As the result, the system filters out the clean, complex mode 
as depicted for example in Fig.~\ref{fig:MovX} or Fig.~\ref{fig:Rot} and determines its growth rate for the assumed frequency.

\subsection{Numerical Procedure, Solver and Configuration}
\label{subsec:25}

The  periodically forced Navier-Stokes computations in frequency domain can be performed by any Computational Fluid Dynamics (CFD) flow solver. 
We use here an extension of in-house Direct Numerical Simulation (DNS) solver based on a second-order finite-element discretization with Taylor-Hood elements \cite{Taylor1973cf} in penalty formulation.  

\begin{figure}
\begin{centering}
\includegraphics[width=0.45\textwidth]{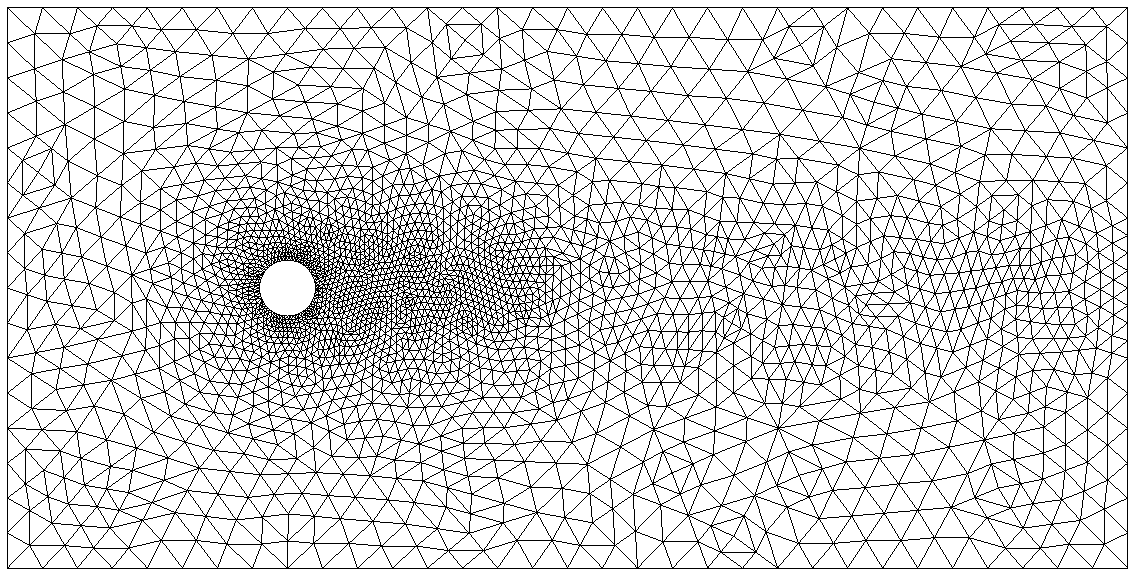}
\hfill{}
\includegraphics[width=0.48\textwidth]{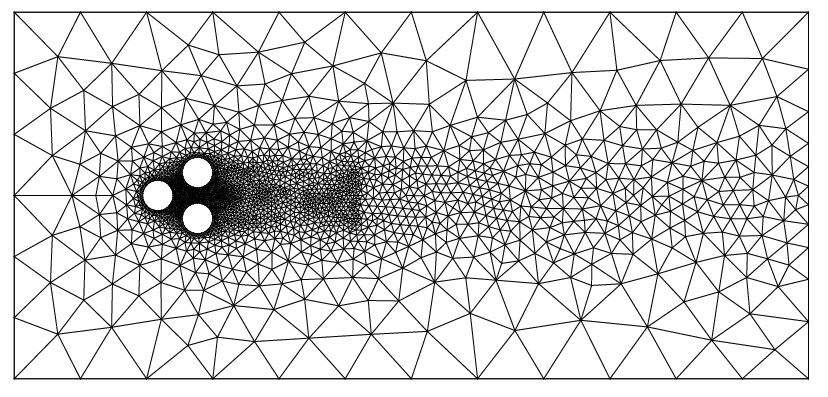}
\par   
\includegraphics[width=0.32\textwidth]{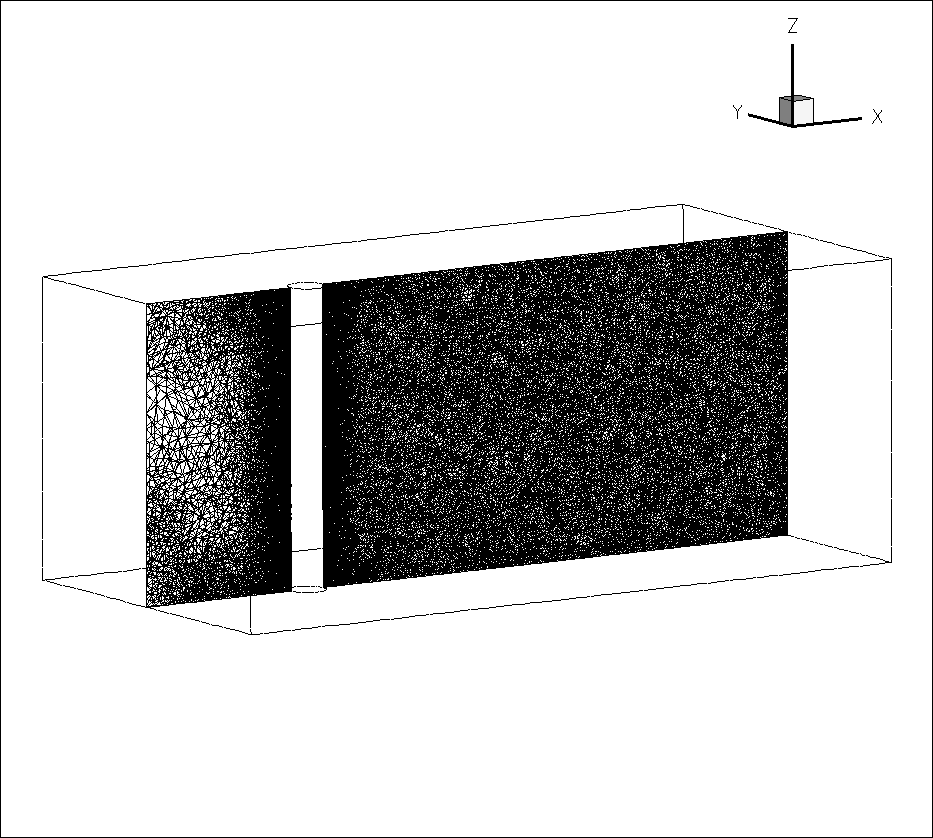} 
\hfill{} 
\includegraphics[width=0.32\textwidth]{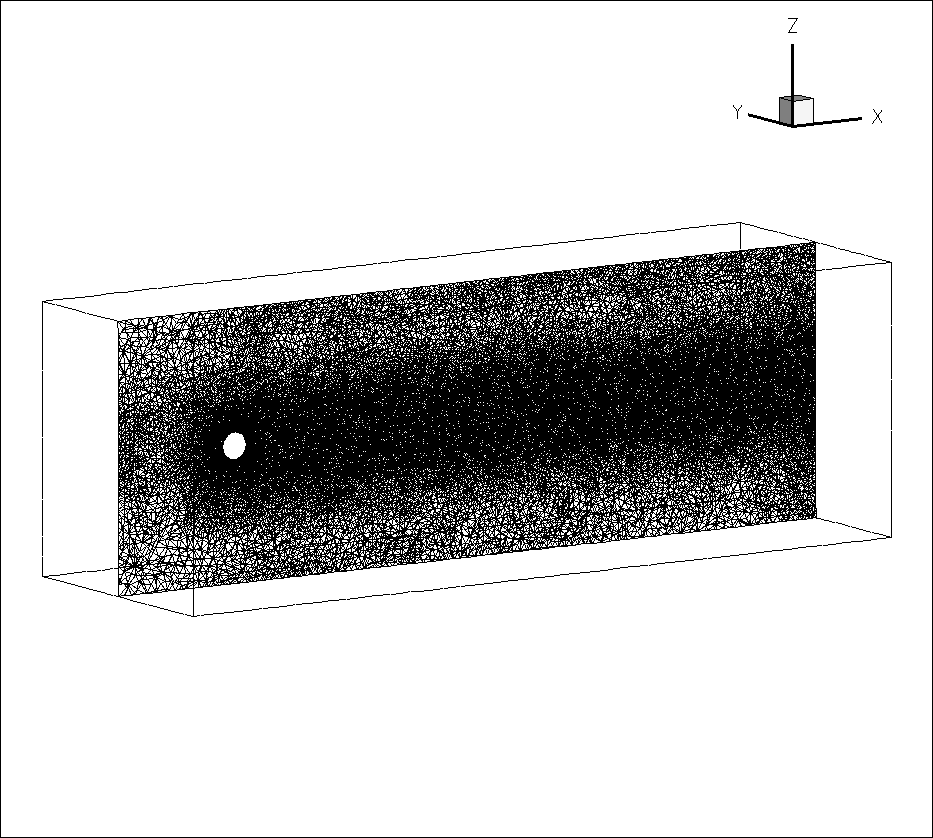} 
\hfill{}  
\includegraphics[width=0.32\textwidth]{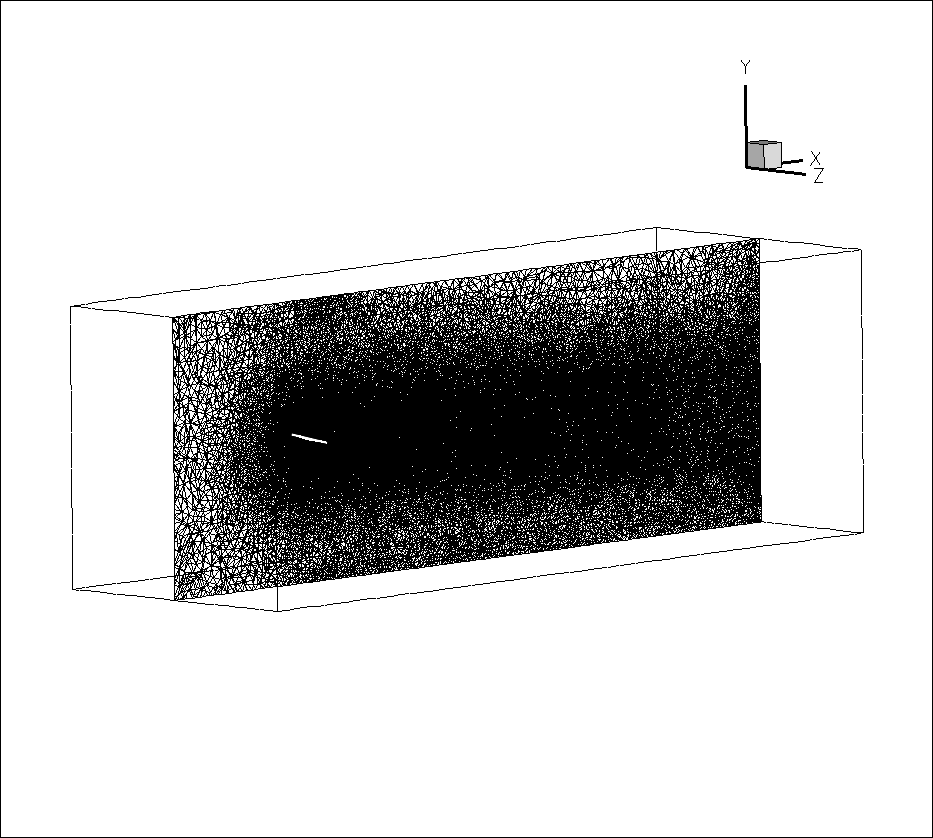}
\par
\end{centering}
\caption{Grids used in 2D and 3D computations.
The top row shows the 2D grids for the cylinder (left) and pinball configuration (right).
The bottom row visualizes the 3D grids  for the wall mounted cylinder (left), the sphere (middle) and the delta wing (right)
in a representative plane.
}
\label{fig:Grid} 
\end{figure}

Following Noack et al. \cite{Noack2003jfm}, 
the computational domain for flow around a circular cylinder, centered at the origin,
extends from $ x=-5 $ to $ x=15 $ and from $ y=-5 $ to $ y=5 $. 
For the fluidic pinball \cite{Noack2017put}
the domain extends from $ x=-6 $ to $ x=20 $ and from $y=-6$ to $y=-6$ 
and is discretized on an unstructured grid with 4225 triangles and 8633 vertices. 
For 3D computation a parallelized version of the code, 
using domain decomposition is employed. 
The grids have respectively  1930678 vertices and  702379 elements for circular cylinder flow, 
960944 vertices and  353310 elements for sphere flow 
and 2401880 vertices and 887845 elements for delta wing flow. 
We use 6-node triangular elements for 2D computations and 10-node tetrahedral ones for 3D simulation.
The number of DOF at each vertex is 2 for 2D and 3 for 3D as the pressure is eliminated with the penalty formulation. The global stability analysis uses the full velocity-pressure formulation. 
Figure \ref{fig:Grid} displays grids employed  for the computations.

\subsection{Steady Navier-Stokes Solution}
\label{subsec:26}

The first step in the solution of \eqref{Eqn:Dist} is to determine the steady base 
solution of the Navier-Stokes equation \eqref{Eqn:NS}. 
At smaller Reynolds numbers, the solution
can be obtained directly with the steady version of \eqref{Eqn:NS}
and an iterative method, like Newton-Raphson.
At higher Reynolds numbers 
the computation can be difficult due to ill-conditioning of the system. 
One remedy is the use 
of Selective Frequency Damping (SFD)~\cite{SFD}. 
Simple unsteady computations with the low order time scheme 
and time step increasing during the computations gives also good approximation of the steady base solution.

\begin{figure}
\begin{centering}
\includegraphics[width=0.6\textwidth]{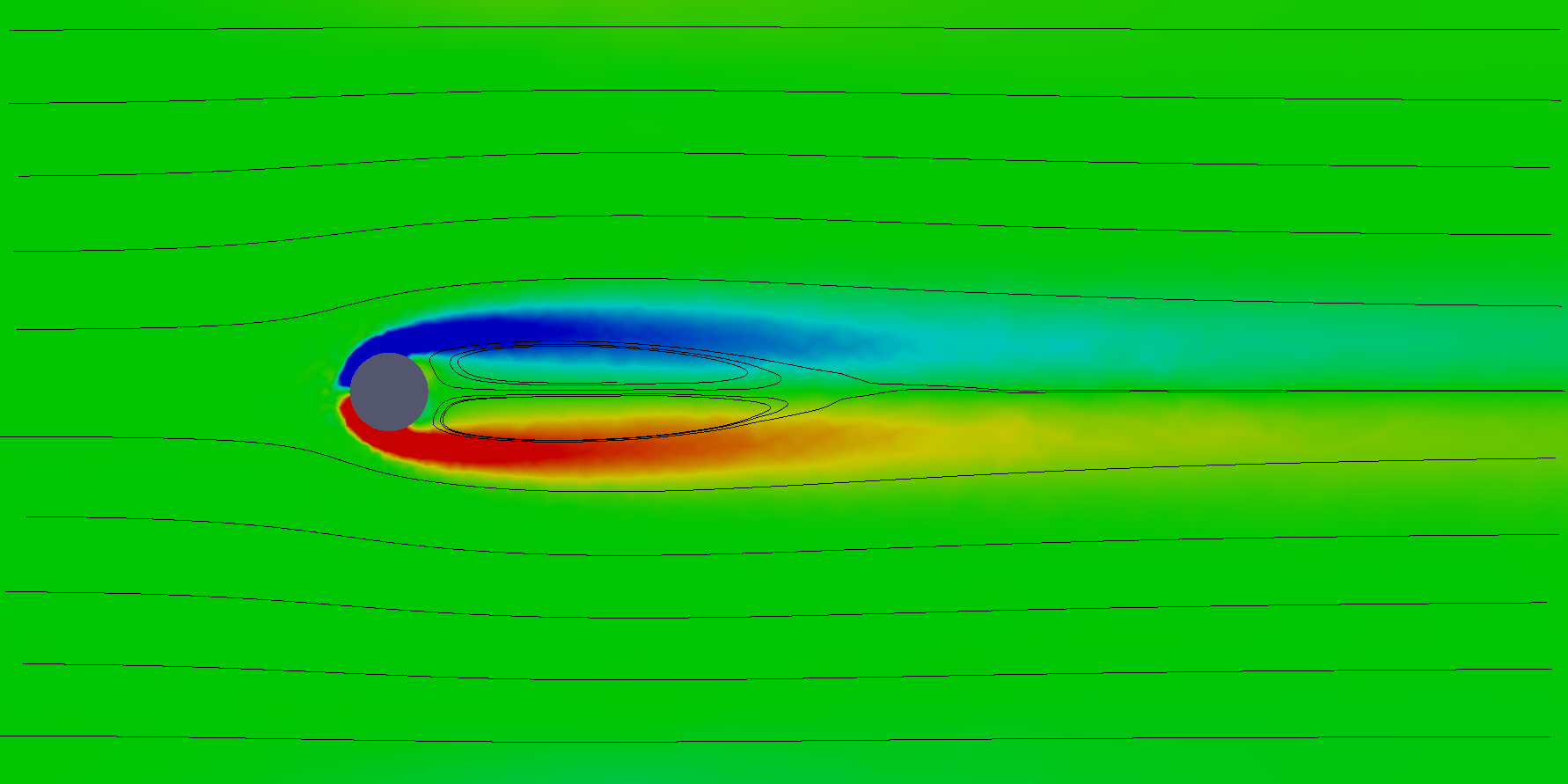}
\par 
\includegraphics[width=0.6\textwidth]{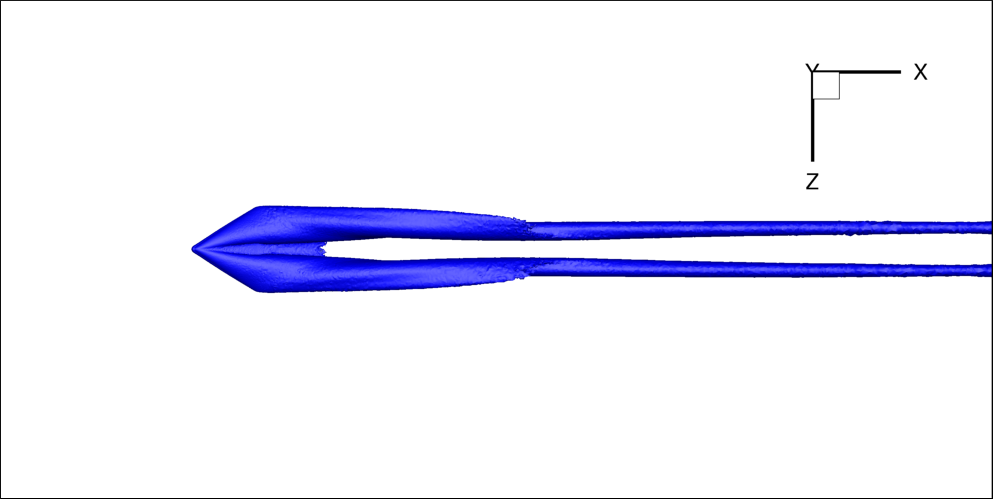}     
\par 
\includegraphics[width=0.6\textwidth]{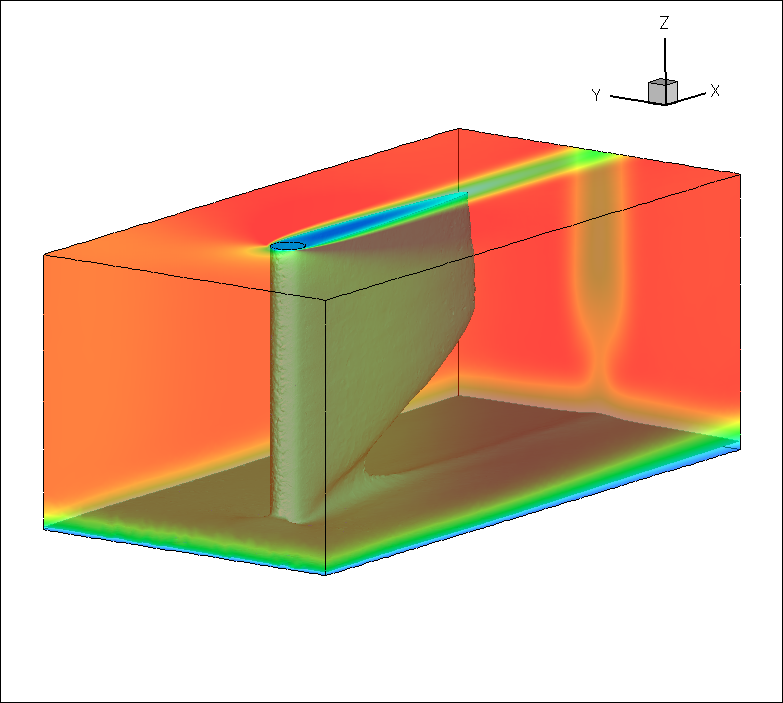}\par
\end{centering}
\caption{Steady Navier-Stokes solutions of the flow around a circular cylinder at $Re=100$ (top, vorticity field),
a non-slender delta wing at $Re=1000$ (middle, velocity isosurface) and a wall-mounted cylinder at $Re=200$ (bottom, velocity isosurface with boundary values).}
\label{fig:Steady_solution}
\end{figure}
In Fig.~\ref{fig:Steady_solution}, 
examples of steady base flows 
are shown for circular cylinder flow, 
the non-slender delta wing and 
the wall-mounted cylinder. 
In all these presented computations, 
we employ only the steady solution of the Navier-Stokes equations 
but the presented methodology is able to handle the flow dynamics
around a mean flow as well.

\section{Two-dimensional Flows}
\label{sec:3}

\subsection{Wake Stabilization by High-frequency Rotation}
\label{subsec:30}
\begin{figure}
\begin{centering}
\includegraphics[width=0.48\textwidth]{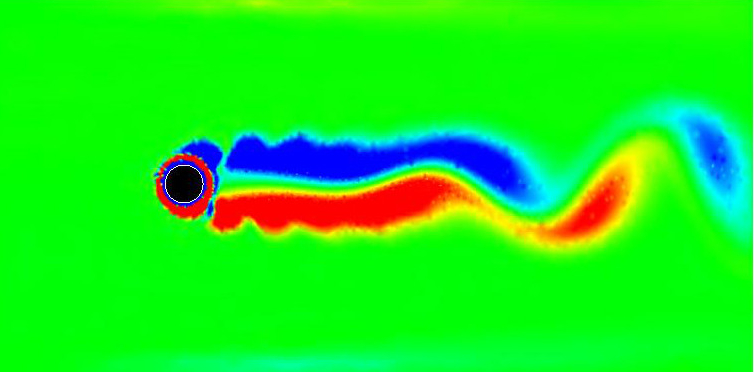} 
\hfill
\includegraphics[width=0.48\textwidth]{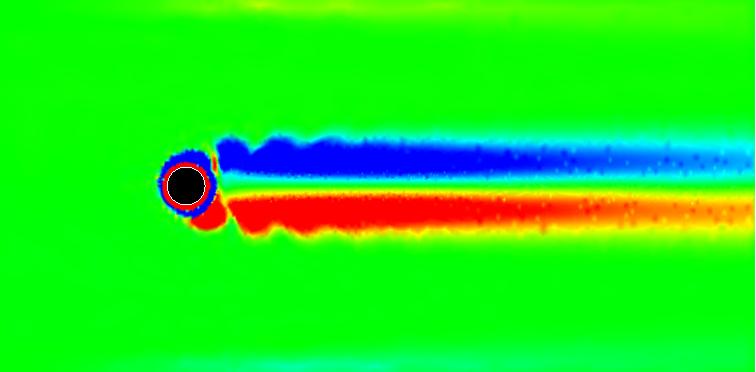}
\par
\includegraphics[width=0.48\textwidth]{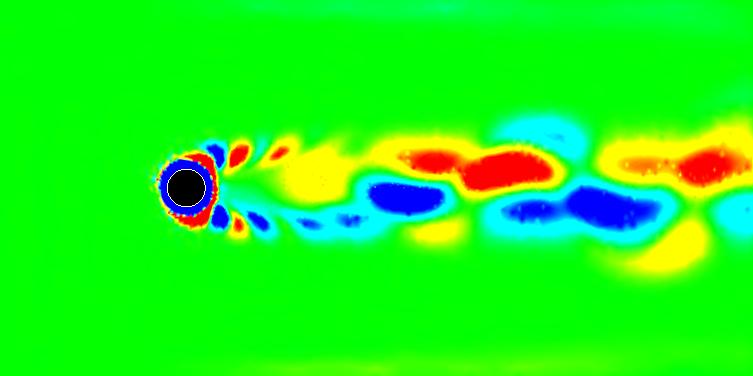}
\hfill
\includegraphics[width=0.48\textwidth]{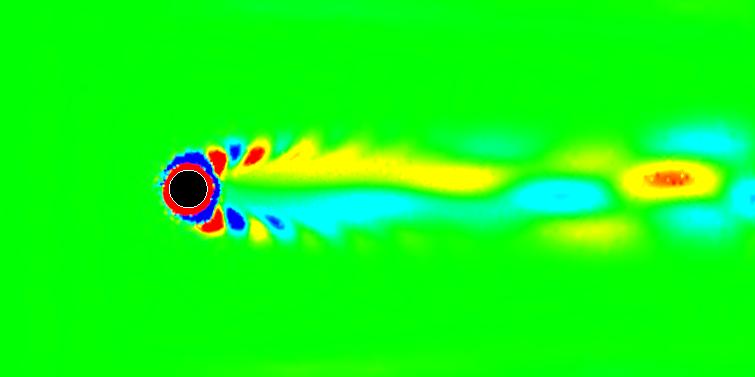}
\par
\end{centering}
\caption{\label{fig:Oscillating_cylinder} Wake of a cylinder  with oscillatory rotation.
The left (right) column visualizes the flow at $t=35$ ($t=200$).
The top (bottom) row corresponds to the vorticity of the solution (disturbance).}
\end{figure}
As the motivation and illustration of the proposed methodology, 
we show the numerical experiment in stabilizing the von K\'arm\'an vortex street with the
rotational oscillation of the cylinder at $Re=100$. 
The oscillation of the cylinder is chosen here to be $4.5$ times of the vortex shedding one.
In Fig.~\ref{fig:Oscillating_cylinder},
we show the effect of the applied control. 
The upper part of the figure shows the
flow field at time instants $t=33$ and $t=200$ 
with initial actuation of the shear layer and the final, stabilized flow. 
The lower part shows the disturbance of the same time snapshot. 
The numerical experiments 
reproduce the water channel investigations of ~\cite{Thiria2006jfm}.
With the frequency domain computations prior to the experiment there is a good opportunity to anticipate the stabilization scenario. 
The effect of oscillatory actuation of a circular cylinder flow is shown in Fig.~\ref{fig:Rot} 
as the pure shear layer mode. 
What distinguishes this from Fig.~\ref{fig:Oscillating_cylinder} is exclusion of the von K\'arm\'an mode 
and its nonlinear interactions.

\subsection{Cylinder Wake}
\label{subsec:31}
\begin{figure}
\begin{centering}
\includegraphics[width=0.48\textwidth]{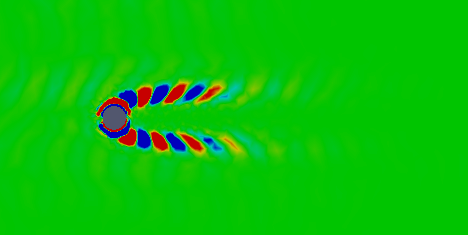} 
\hfill
\includegraphics[width=0.48\textwidth]{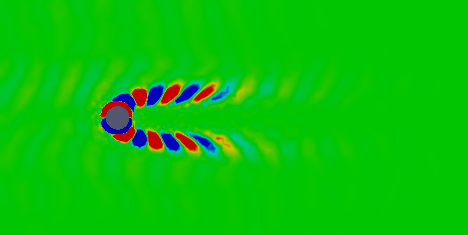}
\par
\includegraphics[width=0.48\textwidth]{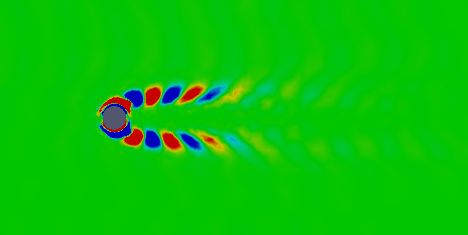}
\hfill
\includegraphics[width=0.48\textwidth]{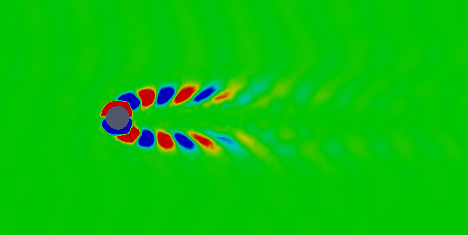}
\hfill
\includegraphics[width=0.48\textwidth]{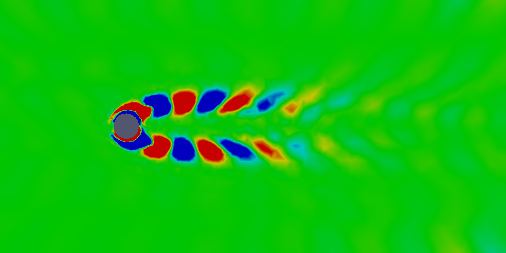}
\hfill
\includegraphics[width=0.48\textwidth]{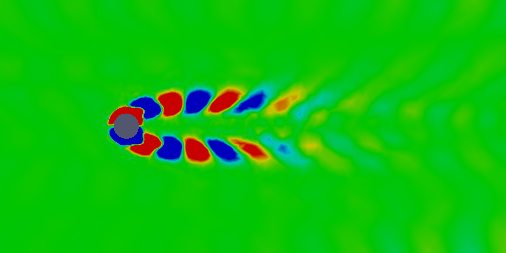}
\par

\includegraphics[width=0.48\textwidth]{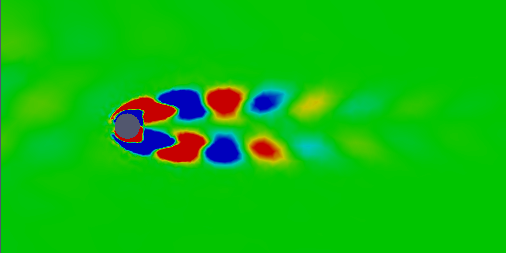}
\hfill
\includegraphics[width=0.48\textwidth]{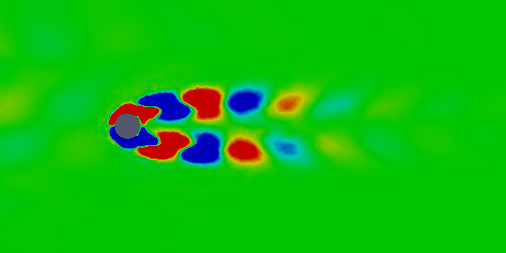}
\par
\includegraphics[width=0.48\textwidth]{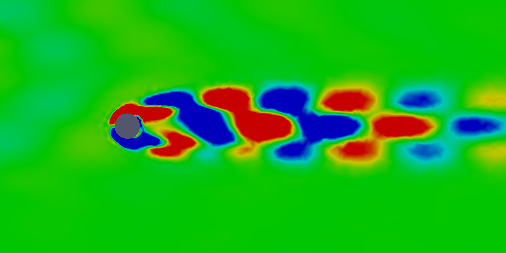}
\hfill
\includegraphics[width=0.48\textwidth]{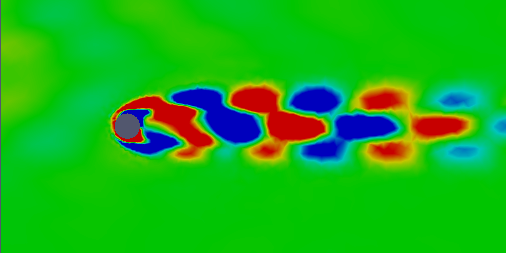}
\par
\includegraphics[width=0.48\textwidth]{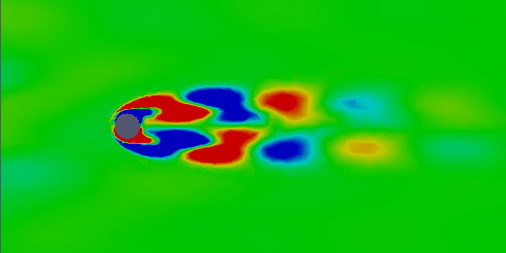}
\hfill
\includegraphics[width=0.48\textwidth]{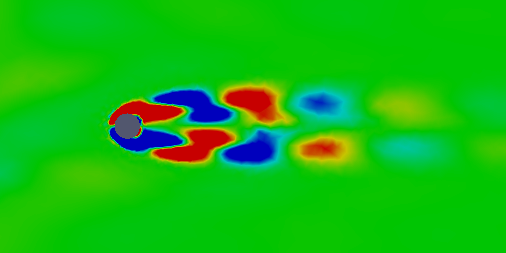}
\par

\end{centering}
\caption{Modes for the circular cylinder wake at $Re=100$ and a symmetric disturbance.  
From top to bottom, 
$\lambda_{Im}$ is equal to $4.2$, $3.2$, $2.2$, $1.2$, $0.87$ and $0.6$.
Real (left) and imaginary (right) vorticity fields are shown.}
\label{fig:MovX} 
\end{figure}

\begin{figure}
\begin{centering}
\includegraphics[width=0.48\textwidth]{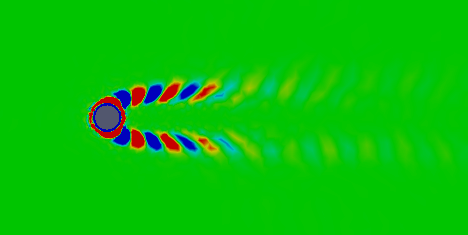} 
\hfill
\includegraphics[width=0.48\textwidth]{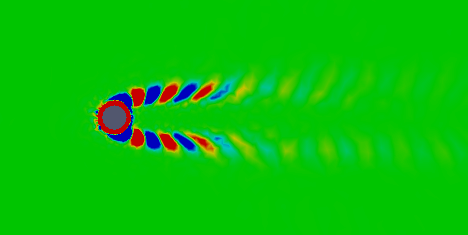}
\par
\includegraphics[width=0.48\textwidth]{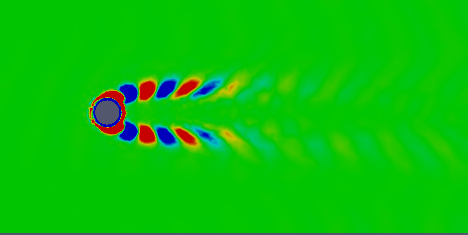}
\hfill
\includegraphics[width=0.48\textwidth]{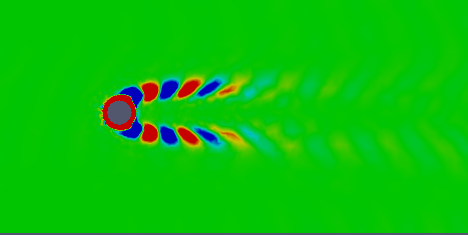}
\hfill
\includegraphics[width=0.48\textwidth]{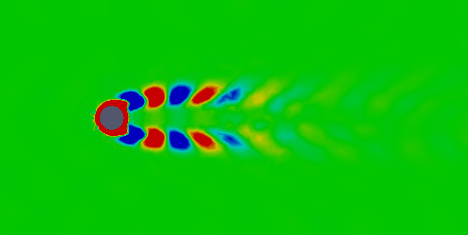}
\hfill
\includegraphics[width=0.48\textwidth]{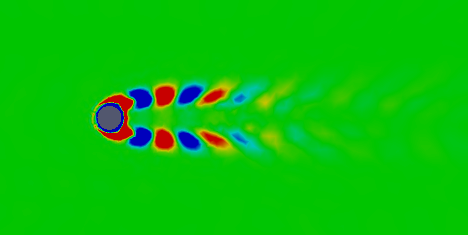}
\par
\includegraphics[width=0.48\textwidth]{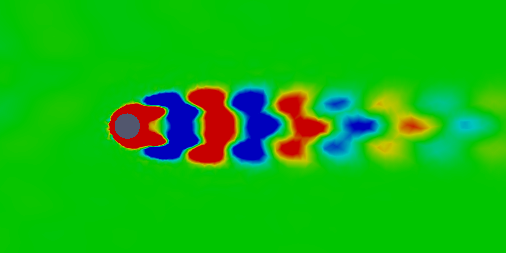}
\hfill
\includegraphics[width=0.48\textwidth]{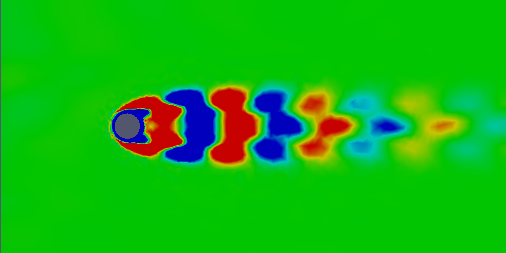}
\par
\includegraphics[width=0.48\textwidth]{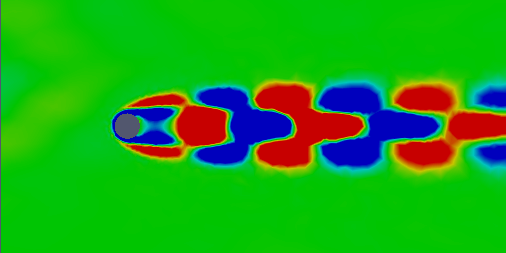}
\hfill
\includegraphics[width=0.48\textwidth]{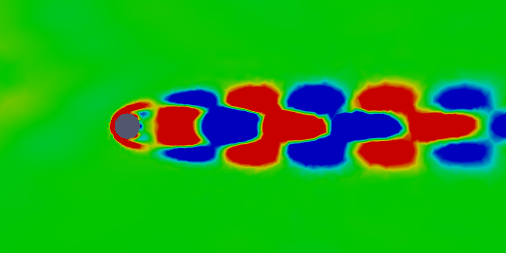}
\par
\includegraphics[width=0.48\textwidth]{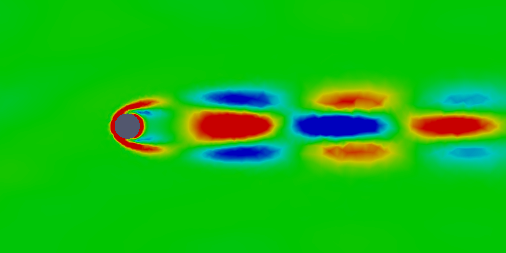}
\hfill
\includegraphics[width=0.48\textwidth]{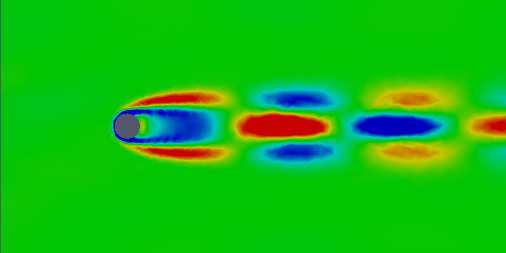}
\par
\end{centering}
\caption{Modes for the circular cylinder wake at $Re=100$ and an oscillatory cylinder rotation.  
From top to bottom, 
$\lambda_{Im}$ is equal to $4.2$, $3.2$, $2.2$, $1.2$, $0.87$ and $0.6$.
Real (left) and imaginary (right) vorticity fields are shown.
}
\label{fig:Rot}
\end{figure}

We consider the unstable steady solution of the circular cylinder flow at $Re=100$.
We show the effect of the symmetric disturbance triggered by a small $x$--axis motion of the cylinder in Fig.~\ref{fig:MovX}  and the antisymmetric one, by a small rotational oscillation of the cylinder in Fig.~\ref{fig:Rot}. 
The imaginary part of the shift, $\lambda_{Im}$ related to the frequency with $\lambda_{Im} = 2 \pi f$ is $4.2$, $3.2$, $2.2$, $1.2$, $0.87$ and $0.6$ respectively. 
The global stability analysis determines the von K\'arm\'an mode frequency $\lambda_{Im}$ for this configuration to be $0.87$. For the symmetric perturbation we obtain the symmetric pattern of the shear flow structures (counter-rotating pairs of vortices, symmetric about x--axis of the flow) for all frequencies except the von K\'arm\'an mode one. Rotation excites co-rotating structures in the shear layer, which merge into the von K\'arm\'an mode at $\lambda_{Im}=0.87$ and form more modes of this type for lower $\lambda_{Im}$.

\begin{figure}

\begin{centering}
\includegraphics[width=0.48\textwidth]{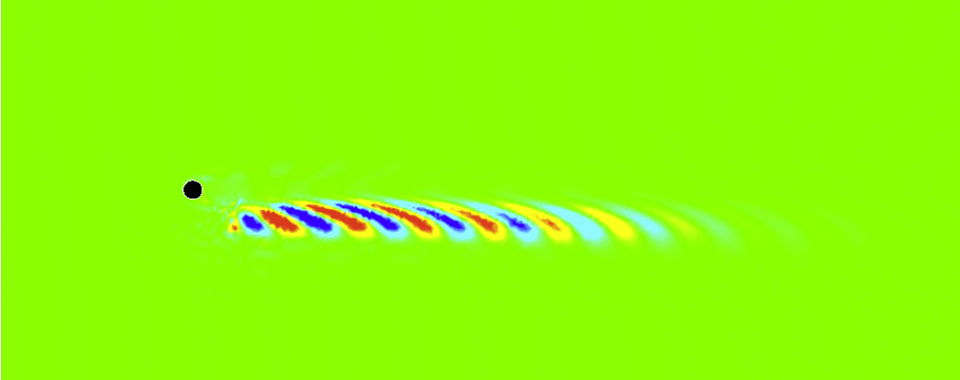}\hfill{}\includegraphics[width=0.48\textwidth]{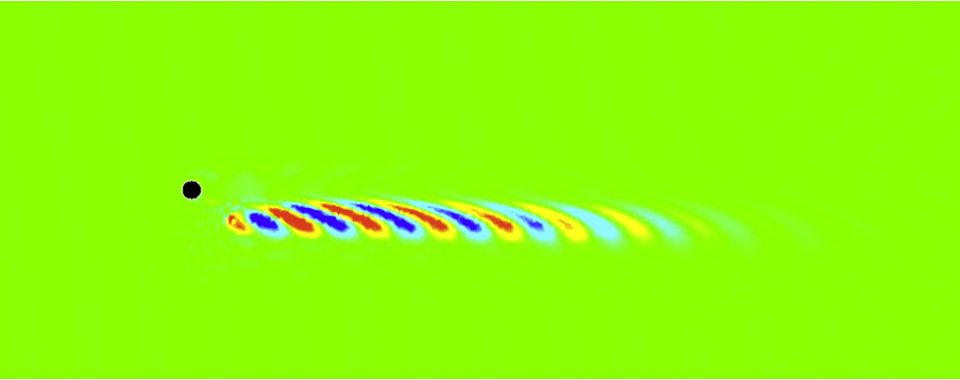}
\par
\end{centering}
\caption{The effect of periodic forcing with a point volume force, $\lambda= 0.13 + i \, 1.87$. Vorticity field shown.}
\label{fig:Point_waves}
\end{figure}

It is also possible to use for actuation the volume force placed in the flow field. 
In Fig.~\ref{fig:Point_waves} the actuation with the point volume force is demonstrated. 
Further experiments with the volume force are shown in Sect.~\ref{subsec:32}.

\subsection{Fluidic Pinball}
\label{subsec:32}

The fluidic pinball is a set of three equal circular cylinders with radius $R$  
placed parallel to each other in a viscous incompressible uniform flow at speed $U_\infty$.
The cylinders can rotate at different speeds
creating a  variety of steady flow solutions and generating a kaleidoscope of vortical structures. 
The configuration is widely used for evaluation of flow controllers~\cite{{Cornejo2017limsi}} 
as this problem is challenging task for control methods comprising several frequency crosstalk mechanisms~\cite{Noack2017put}. 
The centers of the cylinders form an equilateral triangle with side length $3R$, 
symmetrically positioned with respect to the flow. 
The leftmost triangle vertex points upstream, while the rightmost side is orthogonal to the oncoming flow. 
The origin of the Cartesian coordinate system is placed in the middle of the top and bottom cylinder.
The Reynolds number $Re_D$, based on the diameter of the single cylinder is  $Re_D=100$. 

\begin{figure}
\begin{centering}
\includegraphics[width=0.48\textwidth]{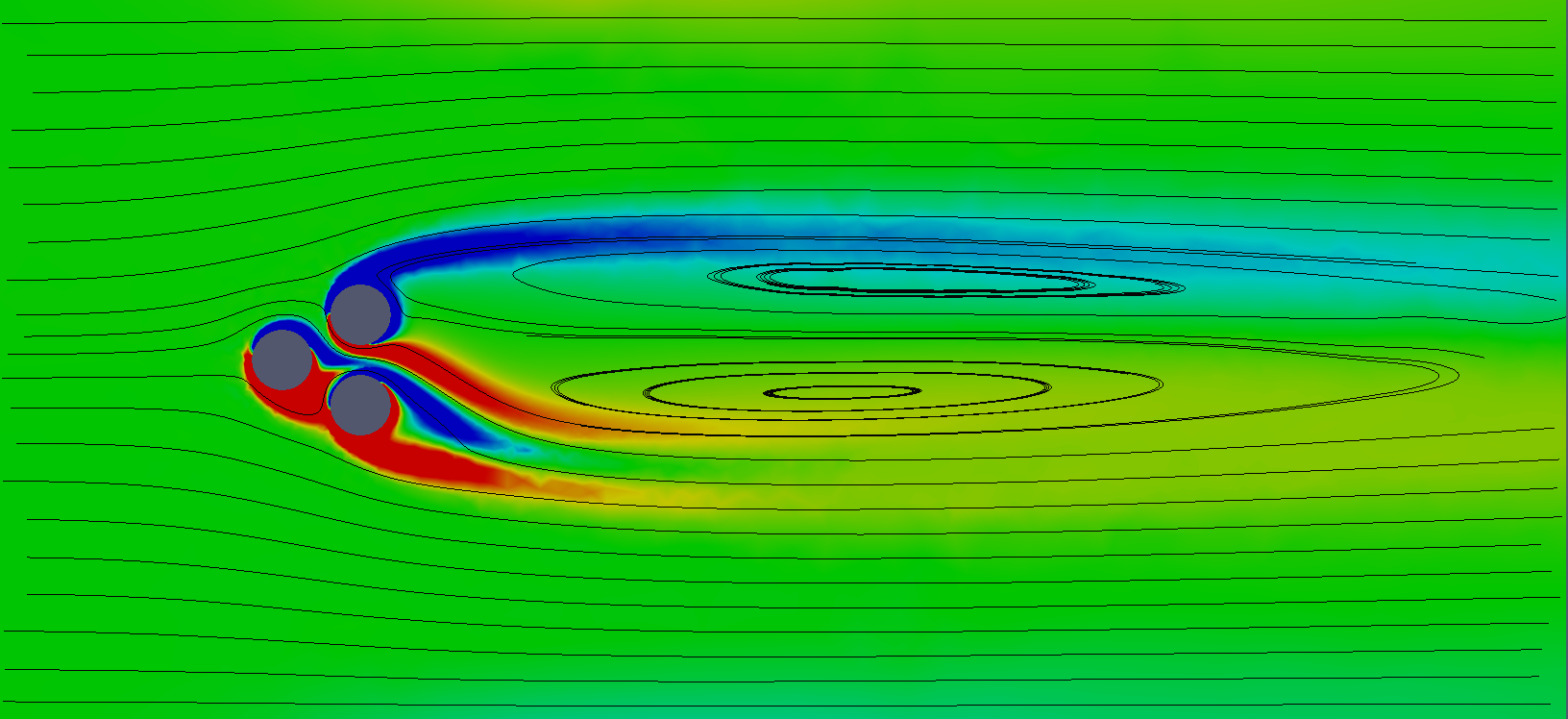}
\par\end{centering}
\caption{Steady Navier-Stokes solution for the fluidic pinball. Vorticity field and flow streamlines are shown.}
\label{fig:Steady_Pinball}
\end{figure}

\begin{figure}
\begin{centering}
\includegraphics[width=0.48\textwidth]{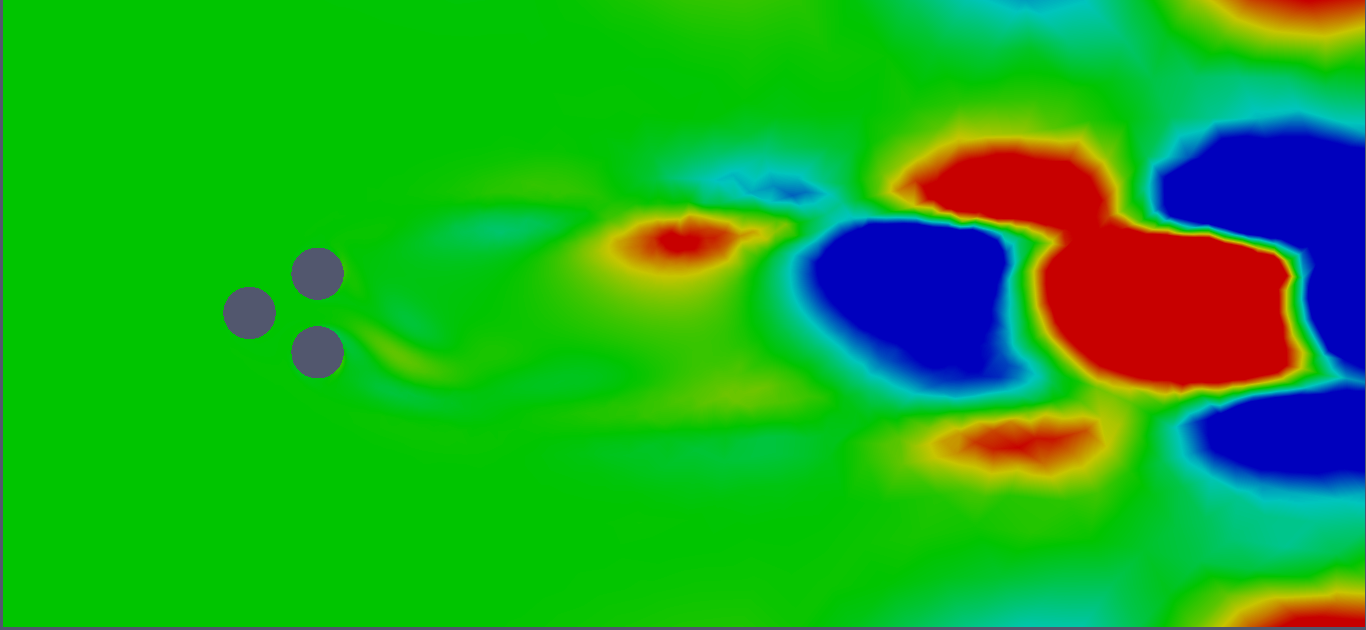}
\hfill{}
\includegraphics[width=0.48\textwidth]{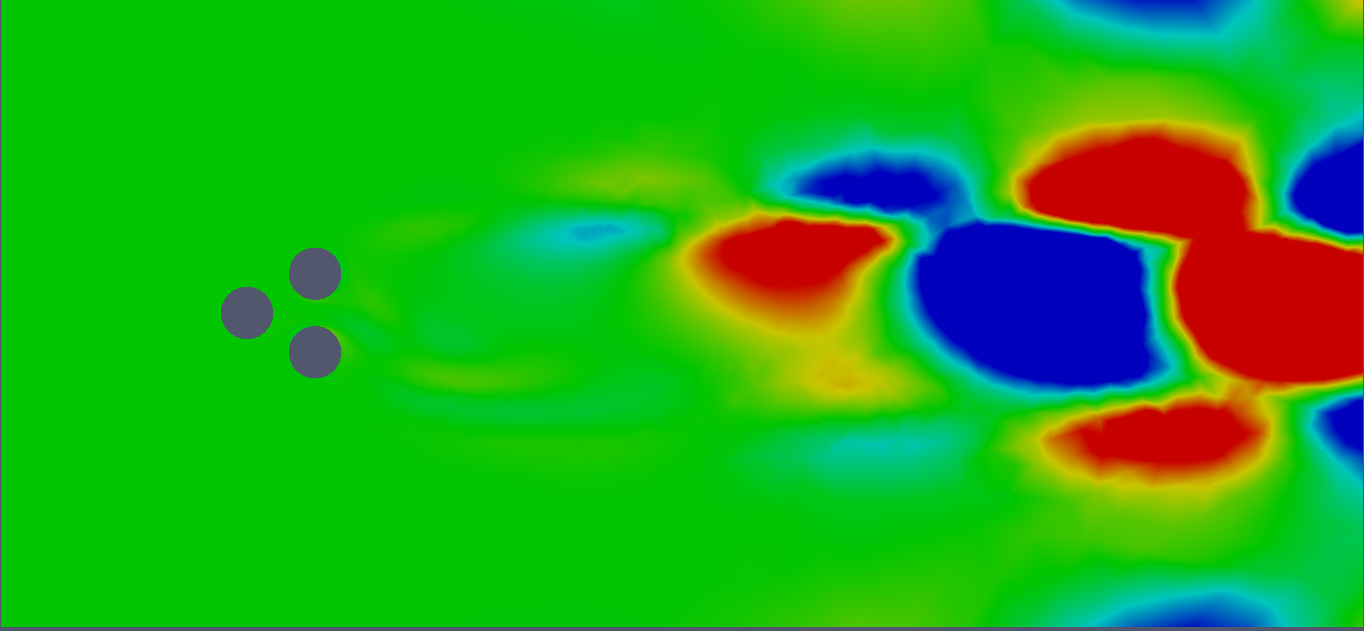}
\par\end{centering}
\caption{Dominating global stability eigenmode for $Re=100$.  $\lambda =   -0.0748 + i  \, 0.466$. Vorticity field shown.}
\label{fig:Karman_Pinball_stability}
\end{figure}

\begin{figure}
\begin{centering}
\includegraphics[width=0.48\textwidth]{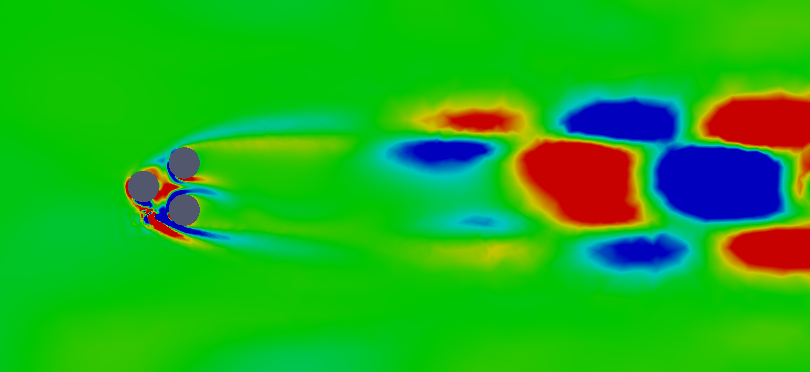}
\hfill{}
\includegraphics[width=0.48\textwidth]{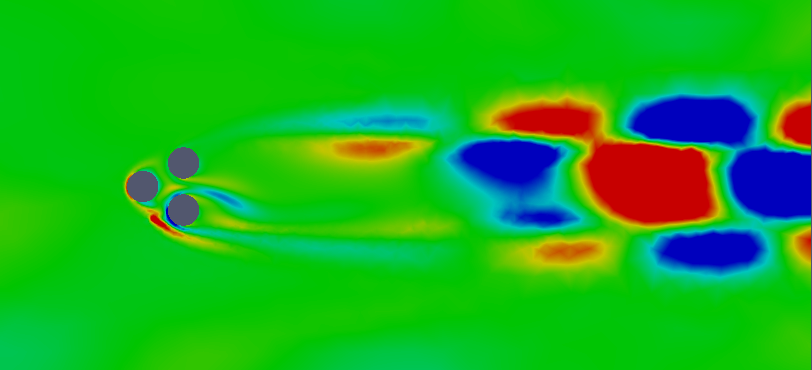}
\par\end{centering}
\caption{ Von K\'arm\'an mode for $\lambda = -0.1 + i 0.45$, actuation at (-1,-1). Vorticity field shown.}
\label{fig:Karman_Pinball_lower}
\end{figure}

\begin{figure}
\begin{centering}
\includegraphics[width=0.48\textwidth]{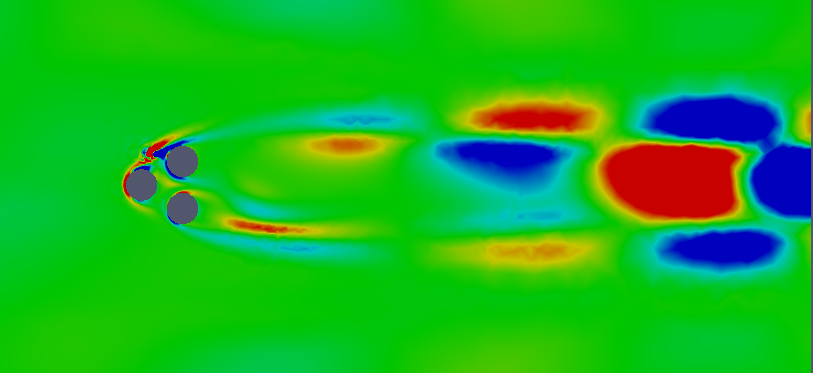}
\hfill{}
\includegraphics[width=0.48\textwidth]{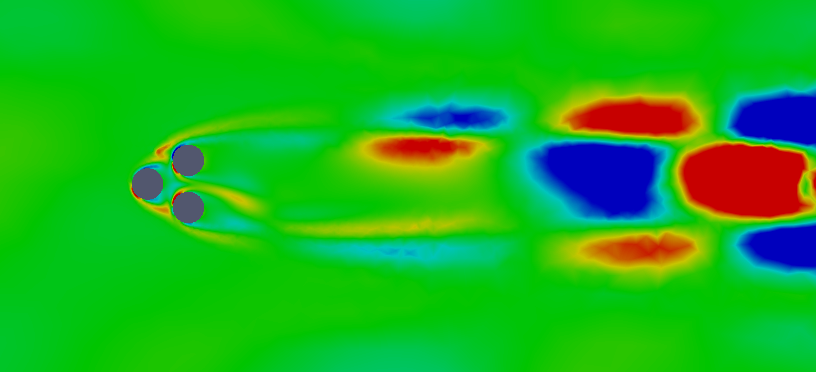}
\par\end{centering}
\caption{\label{fig:Karman_Pinball_uper} Von K\'arm\'an mode for $\lambda = -0.1 + i 0.45$, actuation at (1,1). Vorticity field shown.}
\end{figure}

\begin{figure}
\begin{centering}
\includegraphics[width=0.48\textwidth]{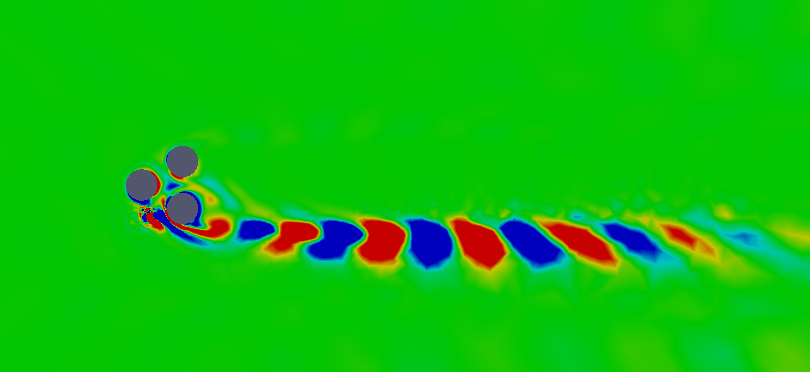}
\hfill{}
\includegraphics[width=0.48\textwidth]{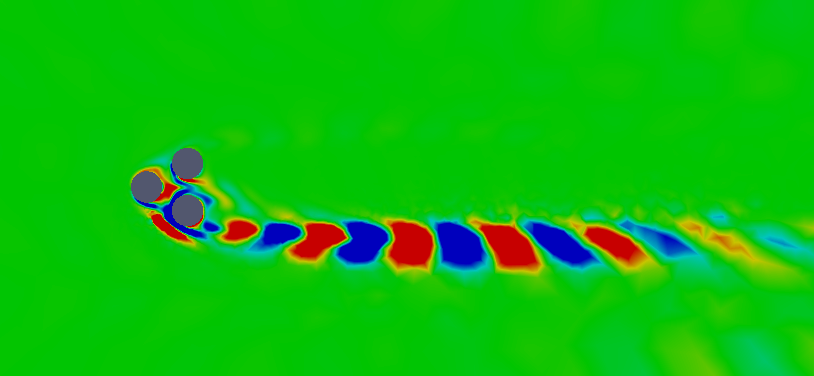}
\par\end{centering}
\caption{\label{fig:Shear_Pinball_lower} Shear layer  mode for $\lambda = 0.2 + i 1.8 $, actuation at (-1,-1). Vorticity field shown. }
\end{figure}

\begin{figure}
\begin{centering}
\includegraphics[width=0.48\textwidth]{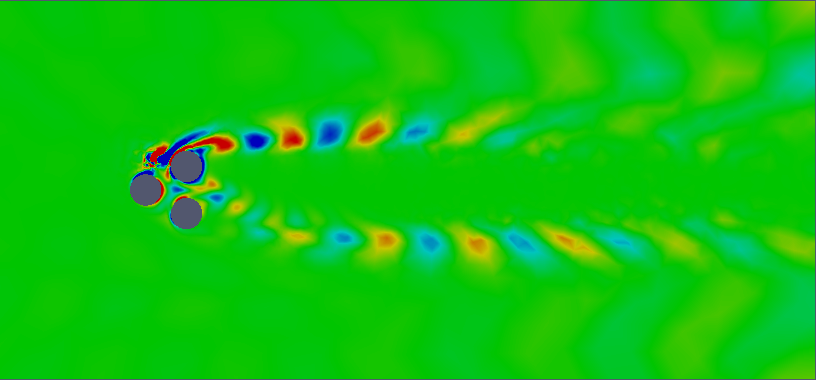}
\hfill{}
\includegraphics[width=0.48\textwidth]{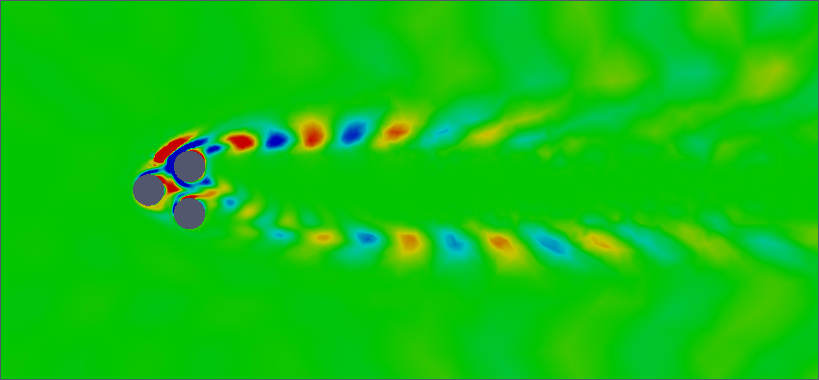}
\par\end{centering}
\caption{\label{fig:Shear_Pinball_upper} Shear layer  mode for $\lambda = 0.2 + i 1.8 $, actuation at (1,1). Vorticity field shown.	}
\end{figure}

To demonstrate the method we compute the steady solution with the upper cylinder rotating counterclockwise, 
the lower one rotating clockwise and the center cylinder also in clockwise direction---all with unit circumferential velocity,
i.e.~is the same as the velocity of oncoming flow. 
The flow-field for this configuration is depicted in Fig.~\ref{fig:Steady_Pinball}. 
The global stability analysis finds the eigenvalue $\lambda =  -0.0748 + i \, 0.466$ 
and the corresponding complex eigenvector depicted in Fig.~\ref{fig:Karman_Pinball_stability}. 
The subsequent two figures 
show modes which are found with the frequency domain approach for $\lambda = -0.1 + i \, 0.45$.
In the Fig.~\ref{fig:Karman_Pinball_lower} the forcing volume force is placed in the point  $(-1,-1)$ while in the Fig.~\ref{fig:Karman_Pinball_uper} at $(-1,1)$. 
Here and in the following examples we use a relatively large value of the actuation to enable visual identification of the actuation placement. The modes closely resemble the
stability eigenmodes and in this case the placement of the actuation is not relevant 
for the resulting flow field.
The effective control of the flow is related to actuation of the different frequencies 
and structures which interact with dominating ones lead us to a desired flow state of 
the fluid. 
To assess this type of actuation we place the actuator, 
represented by the value force at the point $(-1,-1)$ and obtain a clear response of the lower shear layer
as depicted in Fig.~\ref{fig:Shear_Pinball_lower}. 
At the same time the actuator placed at the point $(-1,-1)$ (Fig.~\ref{fig:Shear_Pinball_upper}) splits its effect to both shear layers resulting in less active excitation. 
Both results are intuitive but the computations show the qualitative picture of possible action.

\section{Three-dimensional Flows}
\label{sec:4}
The pronounced complex pair of modes shown in Sect.~\ref{sec:3} is  obtained in relatively few iterations. 
This is in contrast to lengthy time integration and can be
particularly attractive for 3D computations. 
We start the analysis with the academic example of a flow around a wall-mounted circular cylinder and of a sphere.  
In addition, we investigate a  non--slender delta wing having an non-vanishing angle of attack (AoA).

\subsection{Wall Mounted Cylinder}
\label{subsec:41}

\begin{figure}
\begin{centering}
\includegraphics[scale=0.5]{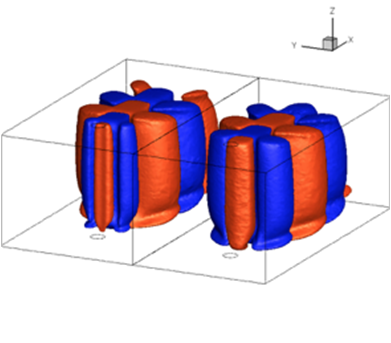}   \hfill{}  
\includegraphics[scale=0.5]{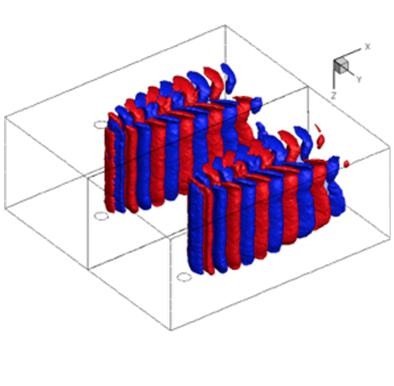}   \par 
\end{centering}
\caption{Flow around a wall-mounted circular cylinder at $Re=300$, $H=8D$.
The left (right) figure shows the real and imaginary part of the von K\'arm\'an  (shear-layer) mode. Both figures include vorticity isosurfaces (red -– negative, blue -- positive).}
\label{fig:3D_Karman_mode}
\end{figure}
\begin{figure}
\begin{centering}
\includegraphics[scale=0.6]{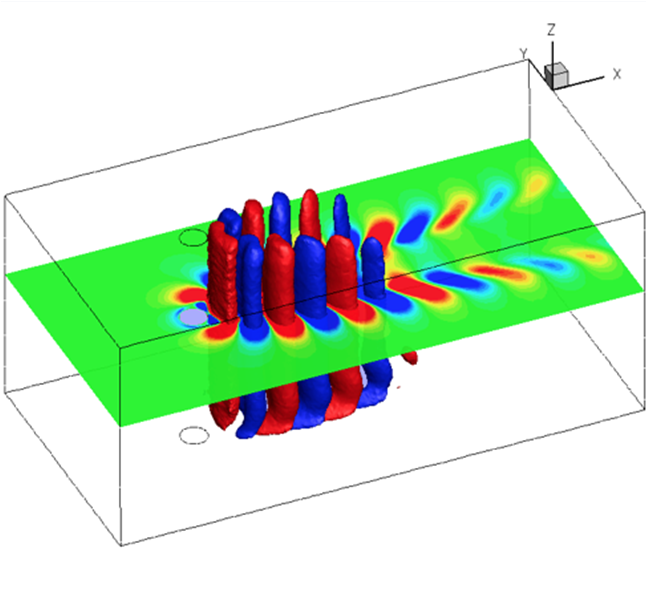}    \par 
\end{centering}
\caption{Close-up view of the real part of the shear-layer mode in Figure \ref{fig:3D_Karman_mode}.}
\label{fig:3D_shear_mode} 
\end{figure}

In Fig.~\ref{fig:3D_Karman_mode} the  3D von K\'arm\'an and shear-layer mode is depicted for $H=8D$ and $Re=300$. 
Like for the two-dimensional cylinder wake,
the real and imaginary part of the modes are shifted in phase.
The bottom wall has a small effect on the dynamics.

In Fig.~\ref{fig:3D_shear_mode} the shear-layer mode is depicted in greater detail.
The antisymmetric flow structures are visualized by iso-surface of $V_1$.
There is mode deformation on the lower surface resulting from no-slip  condition on the wall,
different   from the 2D cylinder flow in Sect.~\ref{sec:2}.

\subsection{Sphere}
\label{subsec:42}

\begin{figure}
\begin{centering}
\includegraphics[width=1.0\textwidth]{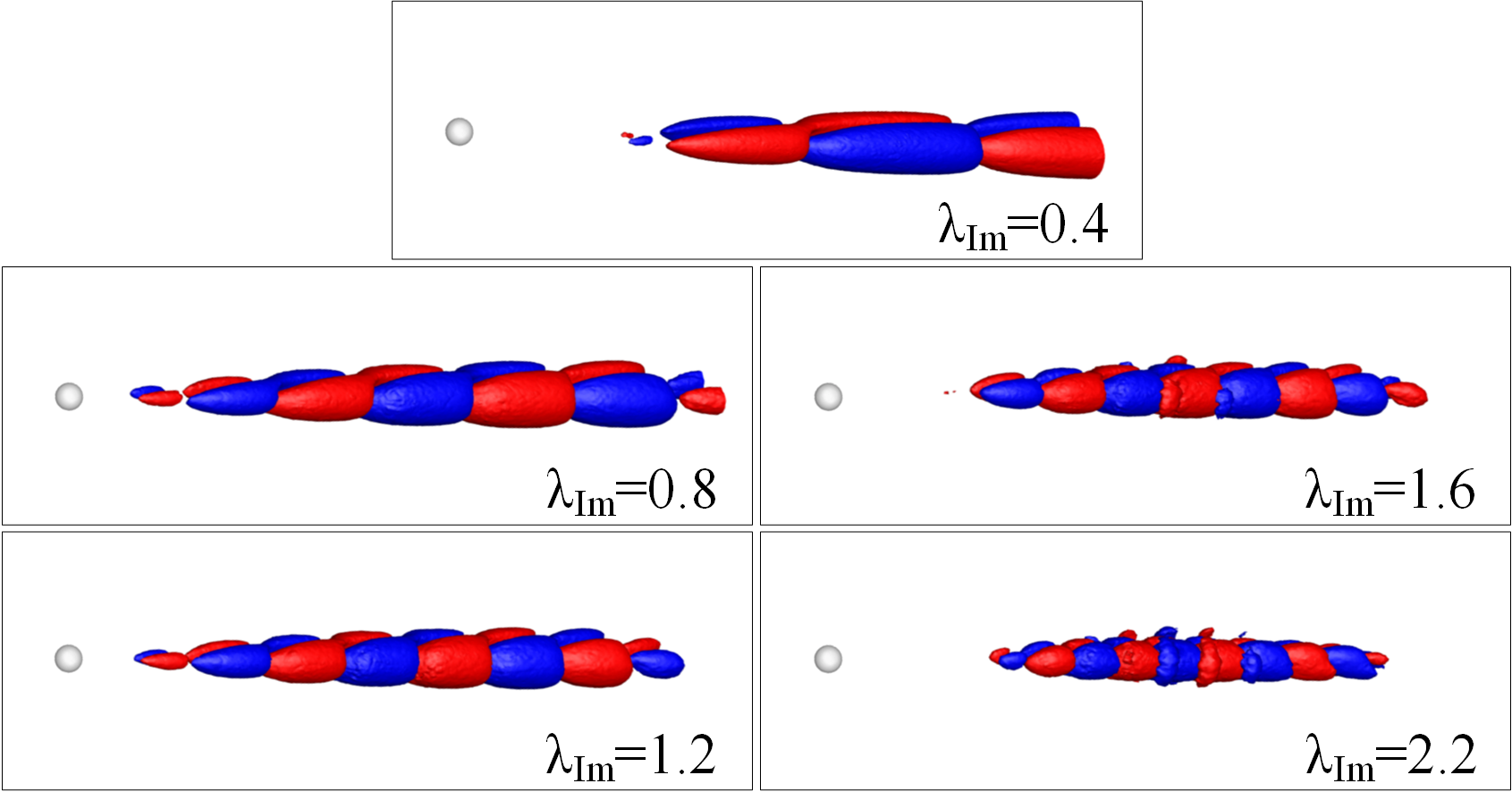}      \par 
\end{centering}
\caption{Sphere modes---Flow response to different frequencies of disturbance. 
Here, $Re = 300$. 
The structures represent iso-surface of the streamwise velocity component (red -– negative, blue -– positive).}
\label{fig:Sphere_flow_different_frequency} 
\end{figure}
The sphere wake is a prototype of many aerodynamic bluff-body flows. 
Here, we present the structures building in the wake of a sphere at $Re=300$. 
In Fig.~\ref{fig:Sphere_flow_different_frequency}, the modes in the wake are present and growing at different
axial location. The closest to the sphere are the modes having frequency closest to the shedding frequency, being the global stability eigenmodes. The frequency is also determining the axial dimension of spirally twisted structures.

\begin{figure}
\begin{centering}
\includegraphics[width=0.48\textwidth]{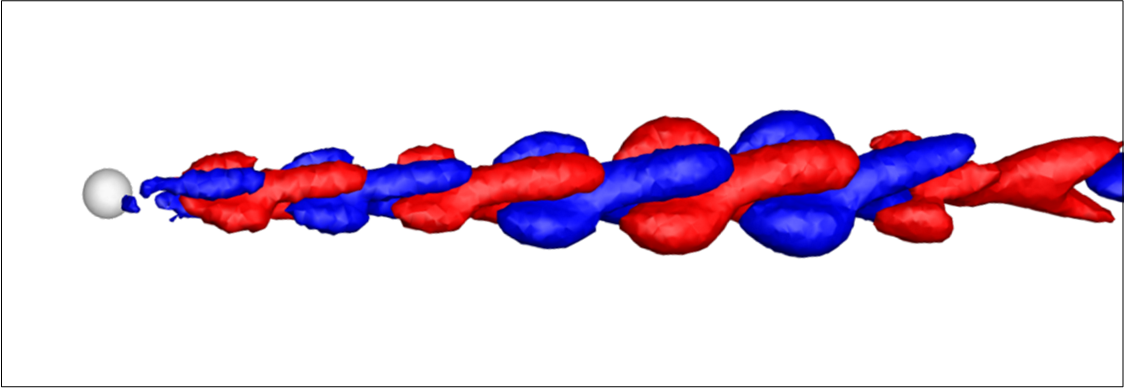}         \hfill{}
\includegraphics[width=0.48\textwidth]{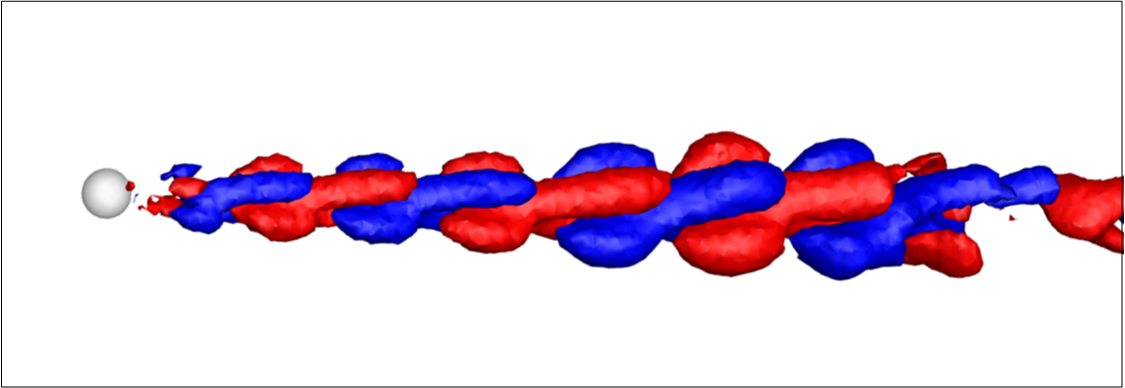}     \par 
\end{centering}
\caption{Sphere wake---DMD mode (real part--left, imaginary part--right) of the transient flow around a sphere at $Re=300$ near the fixed point (isosurfaces of streamwise velocity plotted, red -– negative, blue -– positive).}
\label{fig:Sphere_DMD} 
\end{figure}
\begin{figure}
\begin{centering}
\includegraphics[width=0.48\textwidth]{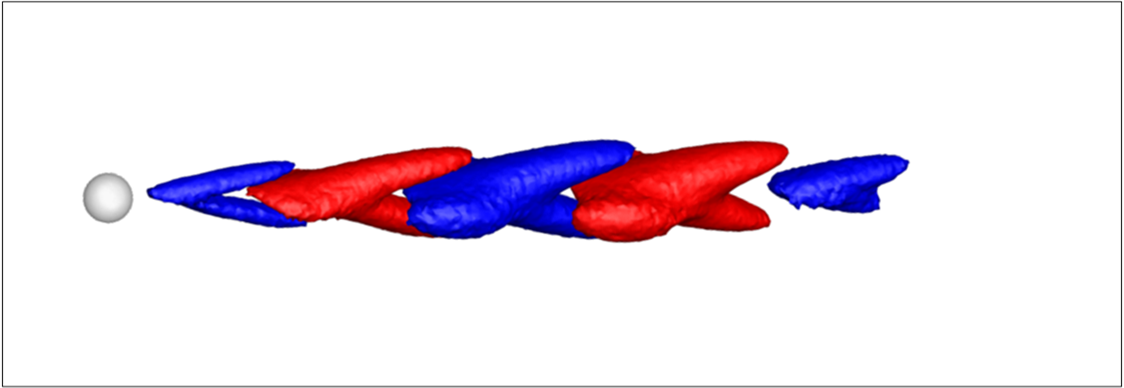}         \hfill{}
\includegraphics[width=0.48\textwidth]{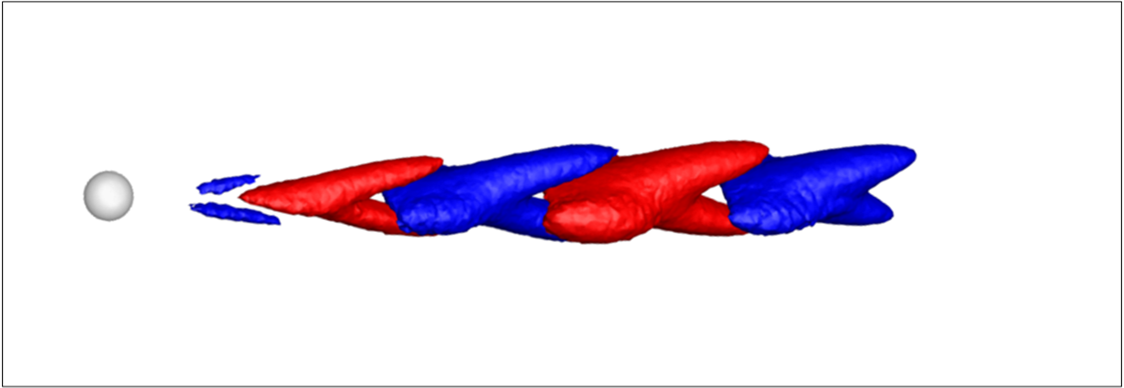}    \par 
\end{centering}
\caption{Sphere wake---Real and imaginary part of the of the frequency domain solution, $Re=300$,  forcing at  $\lambda_{Im} =  0.785$ (isosurfaces of streamwise velocity plotted, red -– negative, blue -– positive).
}
\label{fig:Sphere_shedding}
\end{figure}
In Figures \ref{fig:Sphere_DMD} and \ref{fig:Sphere_shedding}, 
we show the Dynamic Mode Decomposition (DMD) modes for the shedding frequency
and the ones obtained with the frequency domain computations. 
The similarities and differences of both sets of modes are clearly seen.

\subsection{Delta Wing}
\label{subsec:43}

\begin{figure}
\begin{centering}
\includegraphics[width=0.48\textwidth]{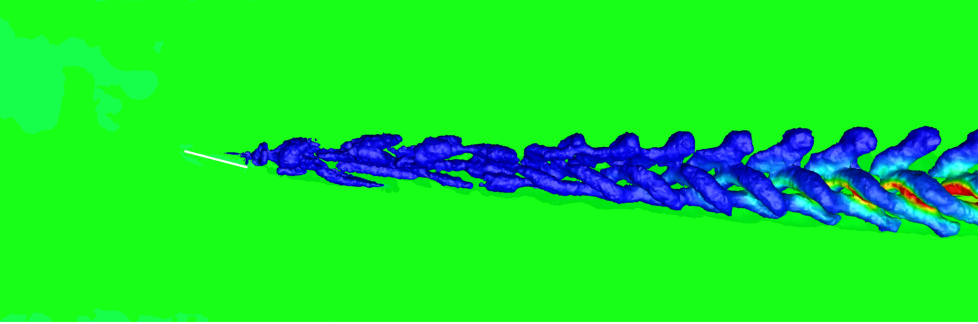}
\hfill{}
\includegraphics[width=0.48\textwidth]{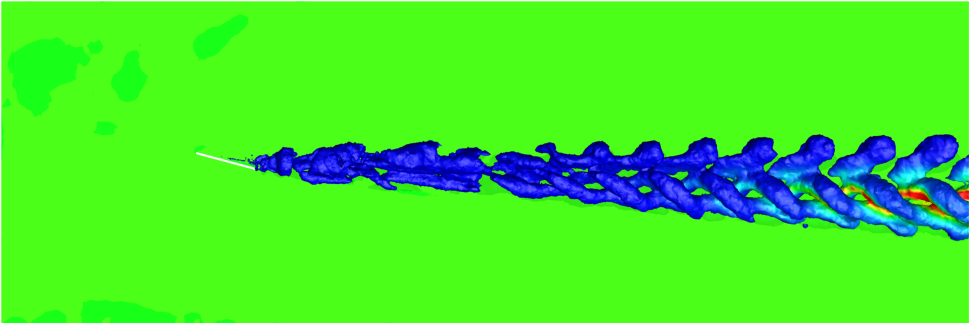}
\hfill{}
\includegraphics[width=0.48\textwidth]{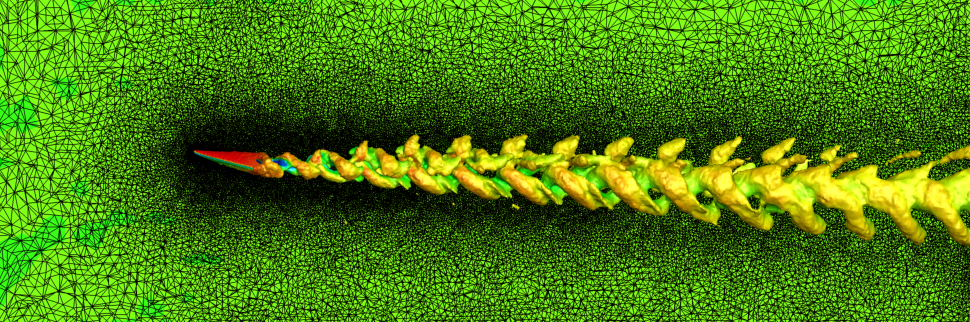}

\end{centering}
\caption{Non-slender delta wing  ($65^\circ$)  at $15^\circ$ AoA, $Re=500$. 
The flow is visualized with $\lambda_2$ iso-surfaces. 
Upper row: real and 
imaginary mode for dominating frequency. 
Lower row: visualization of the fully developed flow computed with the unsteady DNS solver.}
\label{fig:Delta} 
\end{figure}
The non-slender delta wing, having a non-vanishing angle of attack
is close to a real-life configuration and very rich in flow phenomena.
The steady solution, depicted in  Fig.~\ref{fig:Steady_solution}, 
shows a separation region behind the body and 
two vortical structures extending far downstream. 
The unsteady solution, even at relatively low Reynolds number, 
combines sphere-like vortex shedding interplay between vortices shed from the right and left part of the wing
and a swirl around the vortex cores. 
Modes for such a complex flow are of particular interest. 

In Fig.~\ref{fig:Delta}, the real and imaginary part of the dominating mode are shown. 
The flow is obtained for a 
$65^\circ$ triangular wing at $15^\circ$ angle of attack. 
The Reynolds number for this flow is $Re=500$.  
Because of the complexity of the flow, $\lambda_2$ surfaces are depicted. 
In the bottom part of the figure the 
snapshot of an unsteady DNS simulation is depicted also with $\lambda_2$ surfaces. 
The dominant structures are present already in the linear solution of frequency domain formulation.

\begin{figure}
\begin{centering}
\includegraphics[scale=0.2]{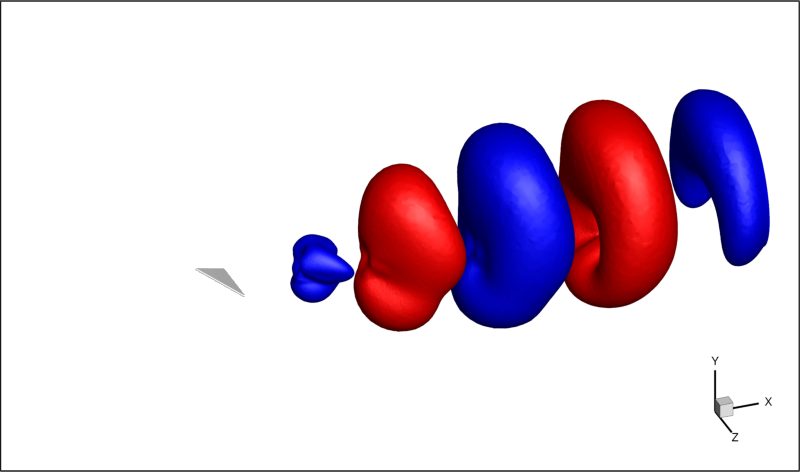}         \hfill{}
\includegraphics[scale=0.2]{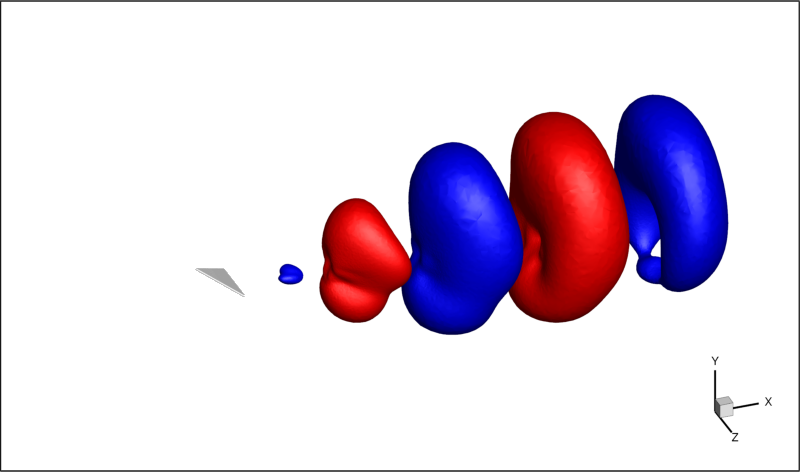}    
\includegraphics[scale=0.2]{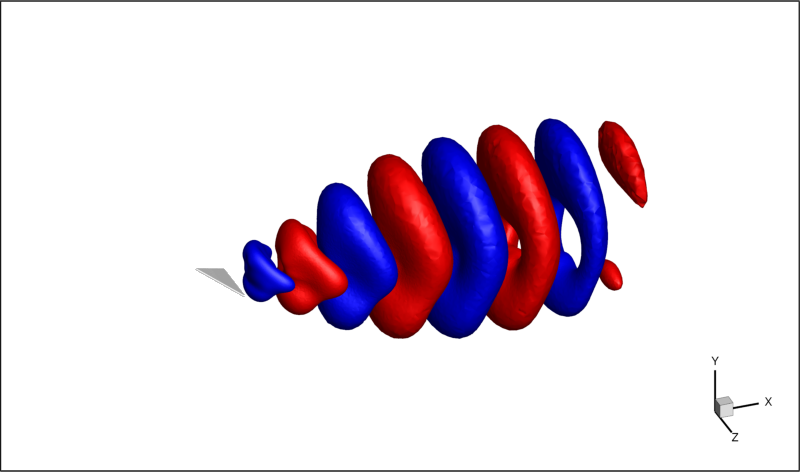}         \hfill{}
\includegraphics[scale=0.2]{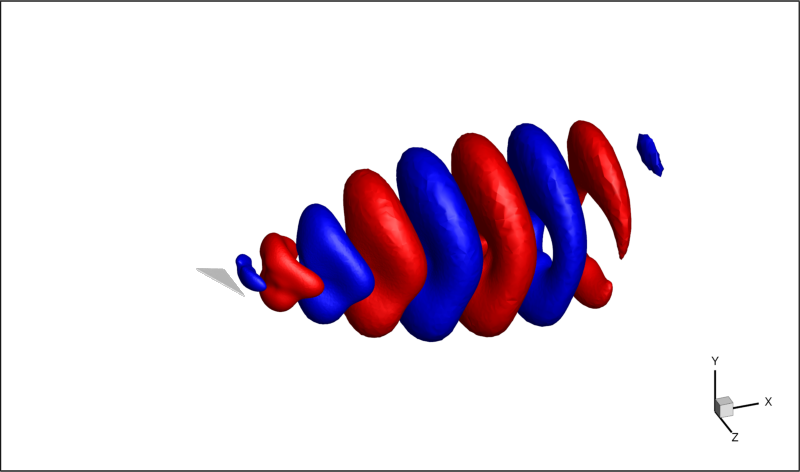} \par 
\end{centering}
\caption{Non-slender delta wing ($65^\circ$), AoA$=30^\circ$, $Re=1000$. 
Complex modes (real part –- left, imaginary part –- right) computed in the frequency domain with forcing frequencies $\lambda_{Im} = 0.75$ (top) and $\lambda_{Im} = 1.5$ (bottom). Isosurfaces of the velocity streamwise component are shown (negative -– red, positive -– blue).}
\label{fig:Delta_wing_65}
\end{figure}
Further figures show the results for the angle of attack of $30^\circ$, 
with iso-lines of 
streamwise velocity for $Re=1000$. 
For both investigated frequencies ($\lambda_{Im}=0.75$ and $\lambda_{Im}=1.5$) the 
modes have shedding character. 

\begin{figure}
\begin{centering}
\includegraphics[scale=0.2]{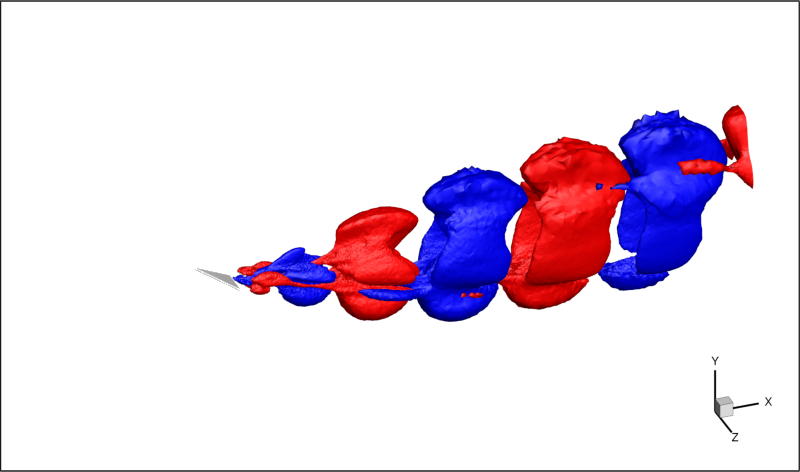}         \hfill{}
\includegraphics[scale=0.2]{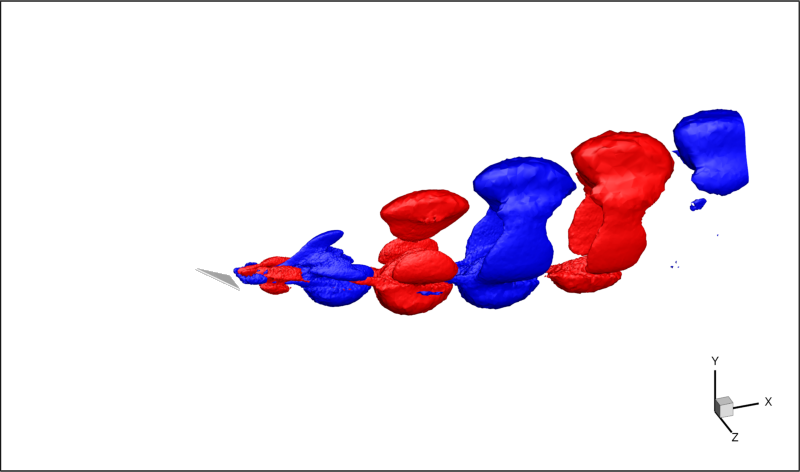}    
\includegraphics[scale=0.2]{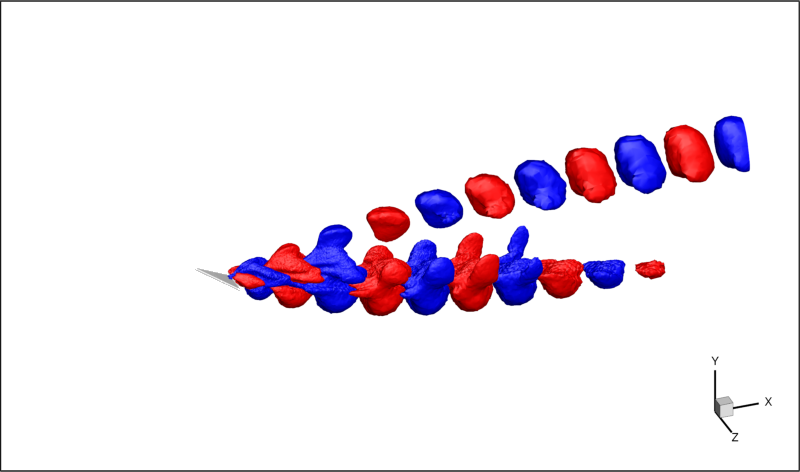}         \hfill{}
\includegraphics[scale=0.2]{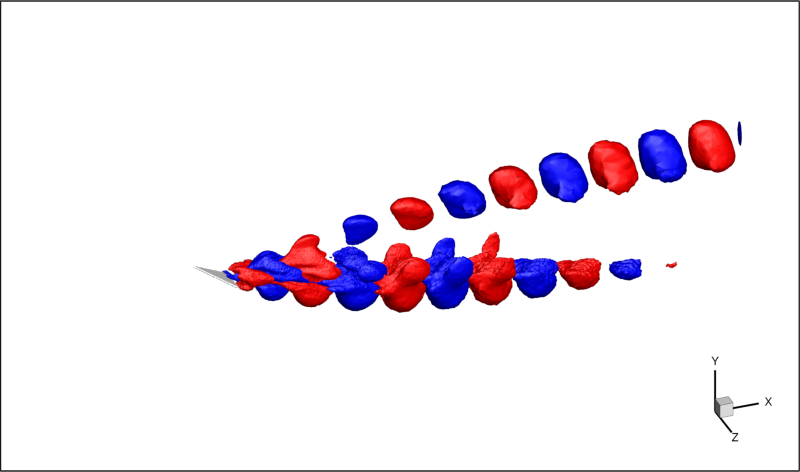}  
\includegraphics[scale=0.2]{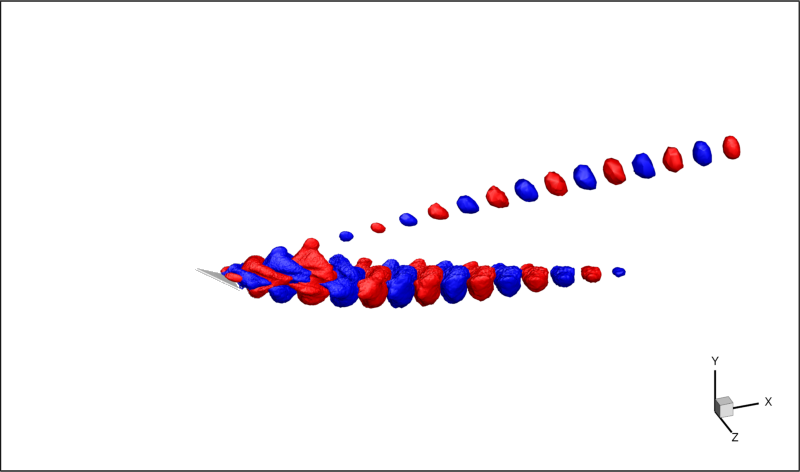}         \hfill{}
\includegraphics[scale=0.2]{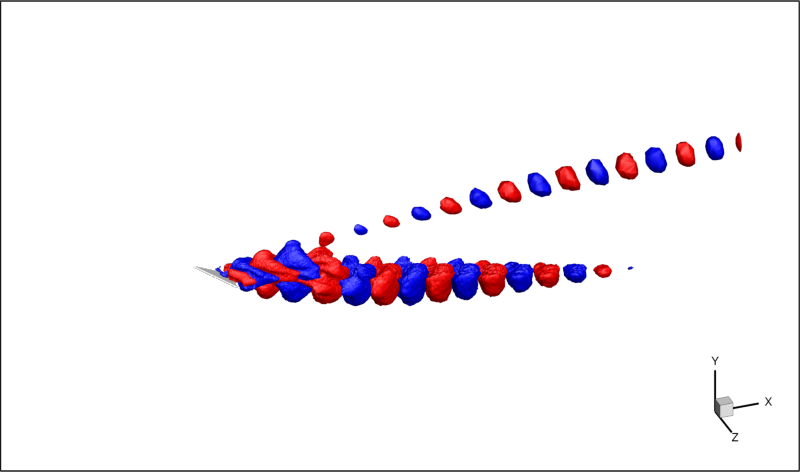}  \par 
\end{centering}
\caption{Delta wing ($35^\circ$), AoA$=30^\circ,$ $Re=1000$.  Complex modes (real part –- left, imaginary part -– right) computed in the frequency domain.  Here  $\lambda_{Im} = 0.75$ (top), $\lambda_{Im}=1.5$ (middle),  $\lambda_{Im}=2.5$ (bottom). Isosurfaces of the velocity streamwise component are shown (negative -– red, positive -– blue).}
\label{fig:Slender_Delta_wing} 
\end{figure}
For a $35^\circ$ wing at the same Reynolds number ($Re=1000$) and angle of attack $30^\circ$ we have 
shedding mode and higher order modes, extending along the steady solution vortex cores (Fig.~\ref{fig:Slender_Delta_wing}).

\section{Nonlinear First-Principle Reduced-Order Modeling of Periodic Wake Actuation}
\label{sec:5}

\begin{figure}
\begin{centering}
\includegraphics[width=0.333\textwidth]{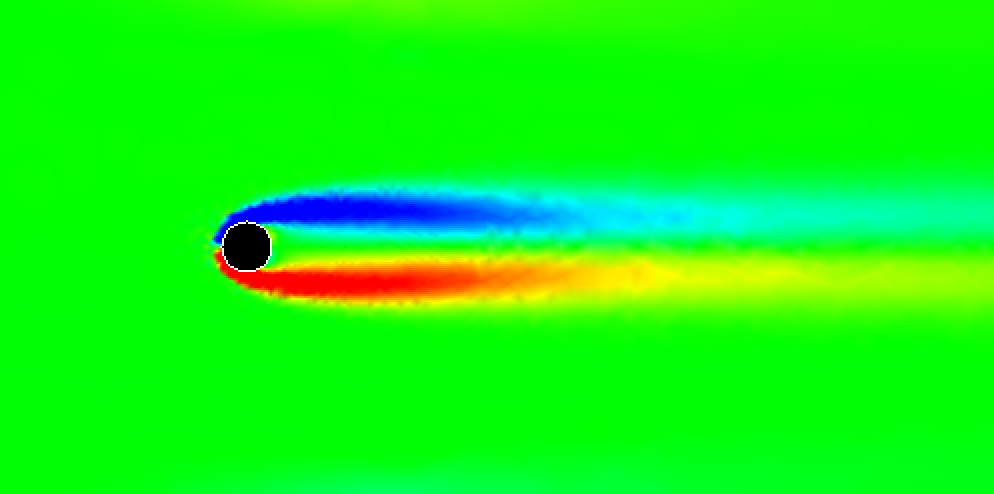}
\includegraphics[width=0.666\textwidth]{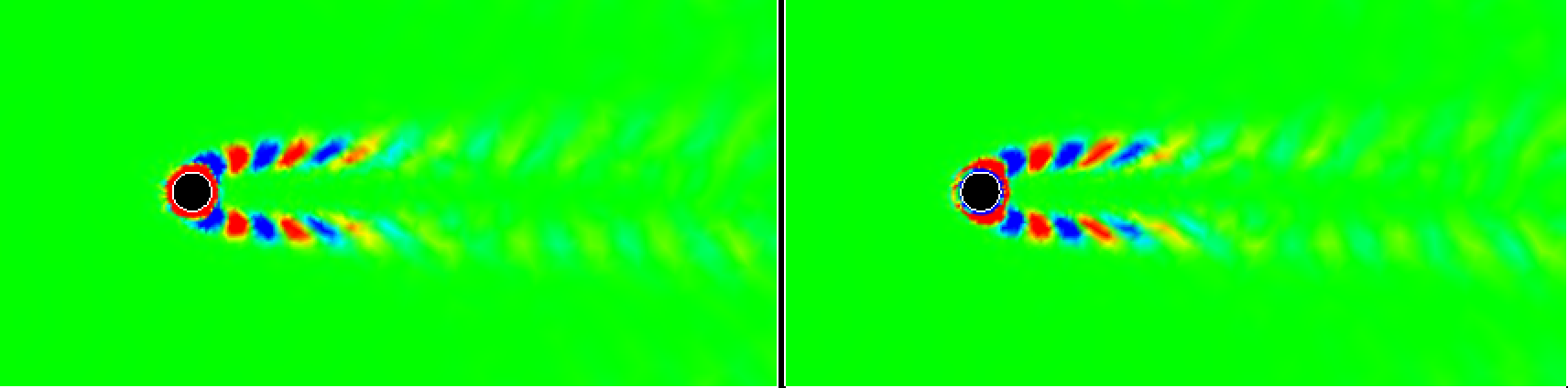}
\includegraphics[width=0.333\textwidth]{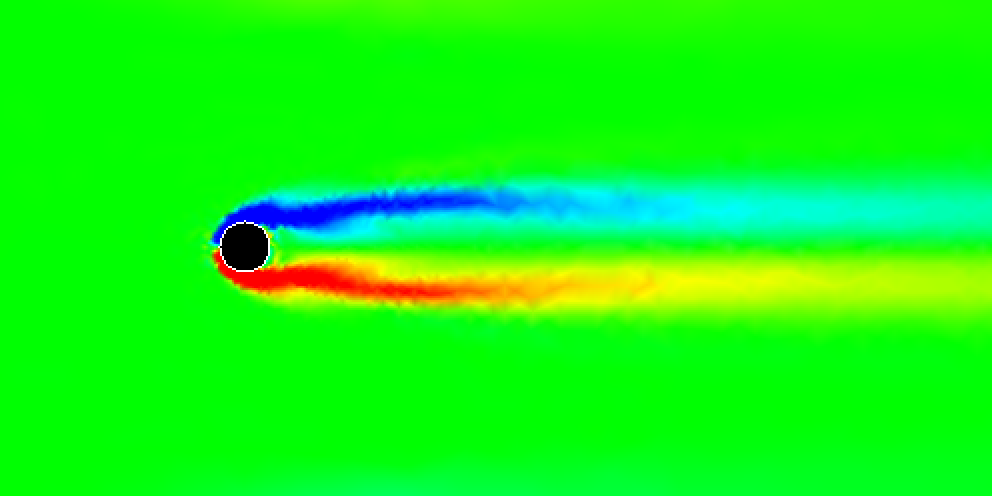}
\includegraphics[width=0.666\textwidth]{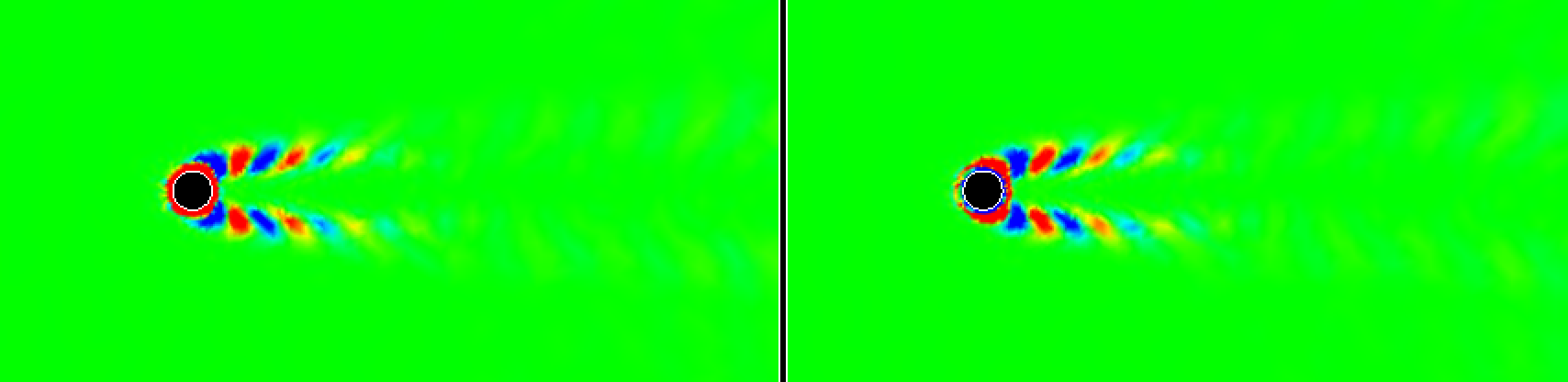}
\includegraphics[width=0.333\textwidth]{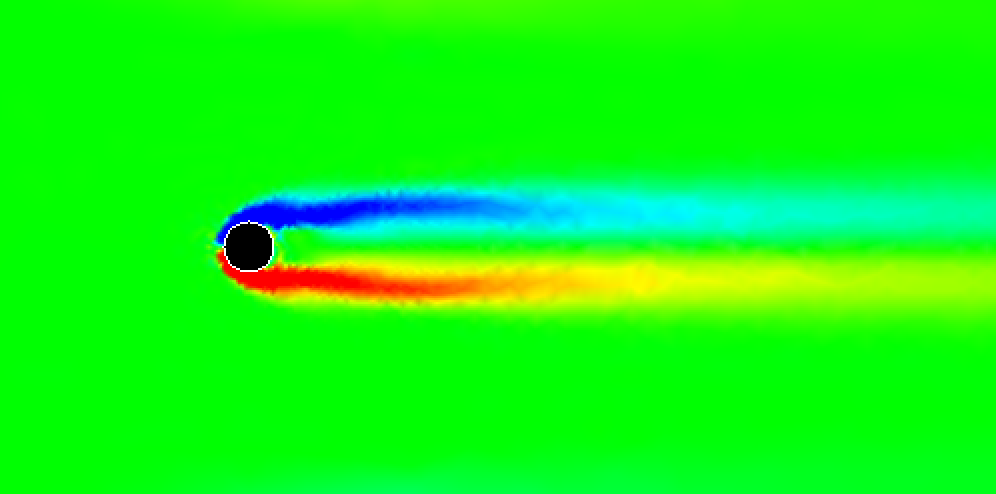}
\includegraphics[width=0.666\textwidth]{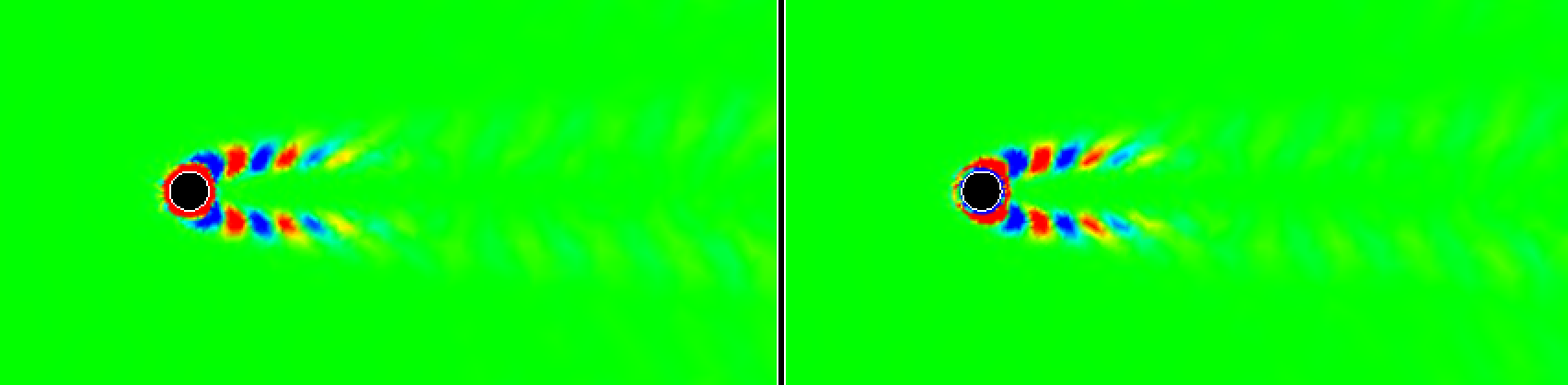}
\includegraphics[width=0.333\textwidth]{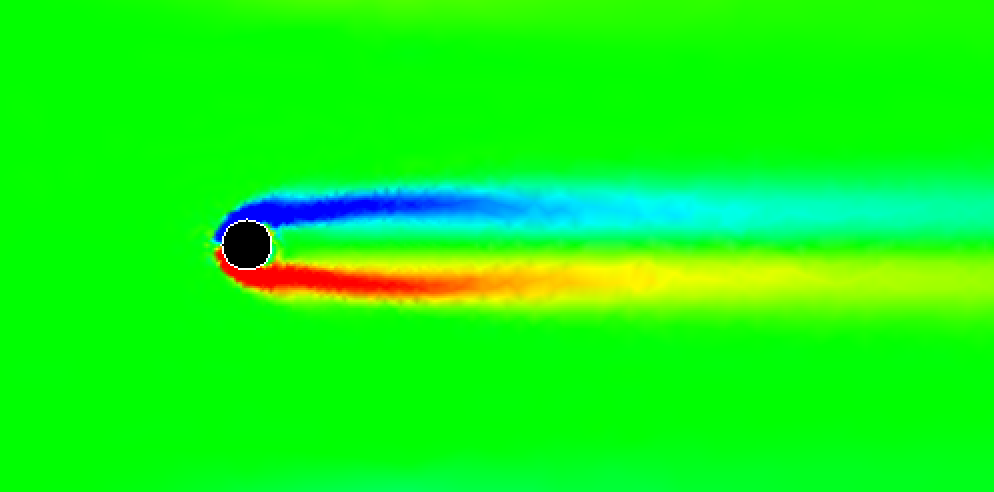}
\includegraphics[width=0.666\textwidth]{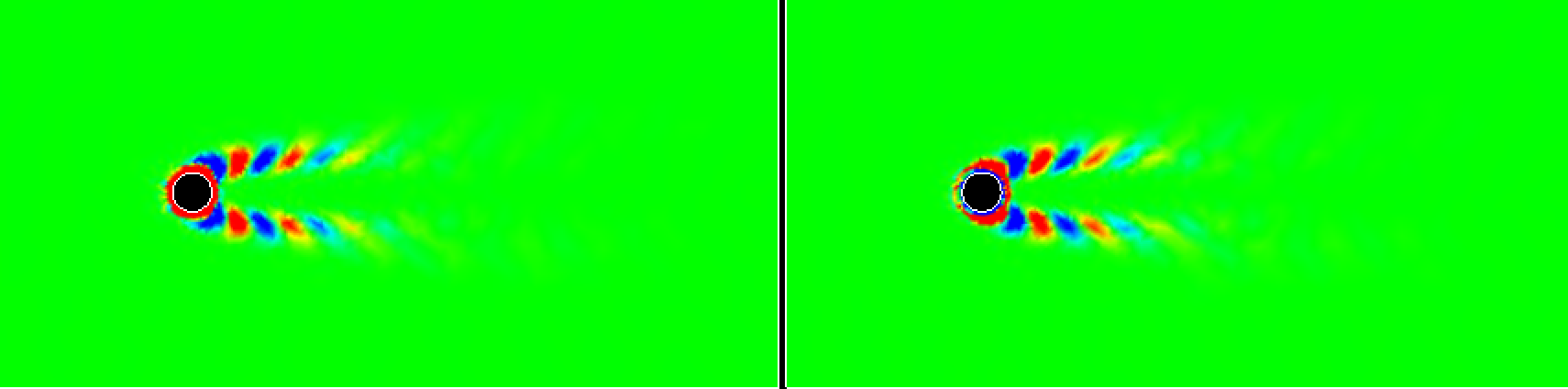}
\end{centering}
\caption{Iterations of the nonlinear model with a physical actuation mode. 
Actuation with oscillatory rotation ($\lambda_{Im}=4.0$) of a circular cylinder at $Re=100$. 
From top to bottom: Vorticity of the mean flow as obtained from Reynolds equation (left)  with corresponding actuation modes (real and imaginary part, right) in iteration 1, 2, 3 and 10.
}
\label{fig:Flow_control} 
\end{figure}
\begin{figure}
\begin{center}
\includegraphics[width=0.8\textwidth]{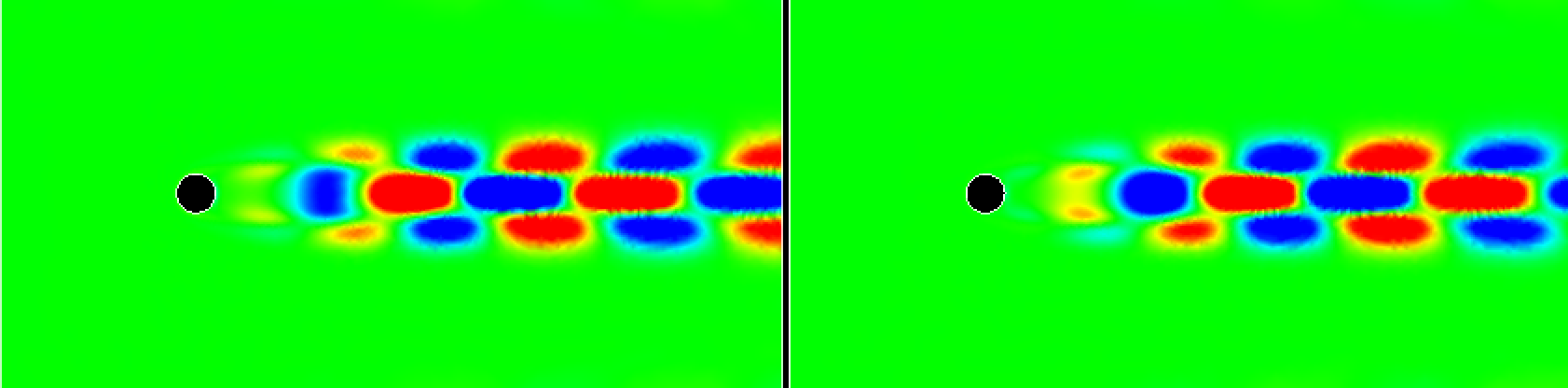}
\includegraphics[width=0.8\textwidth]{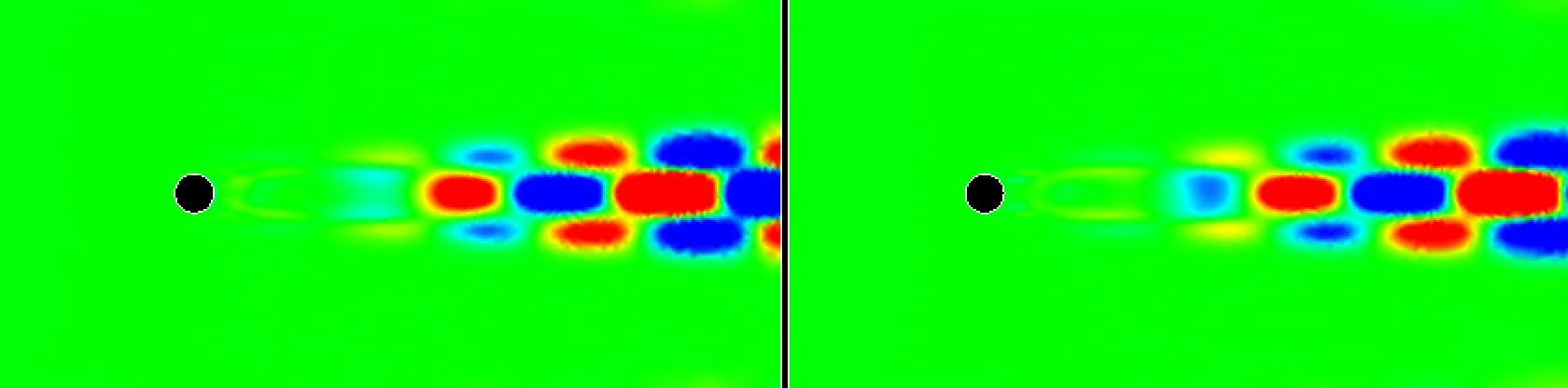}
\includegraphics[width=0.8\textwidth]{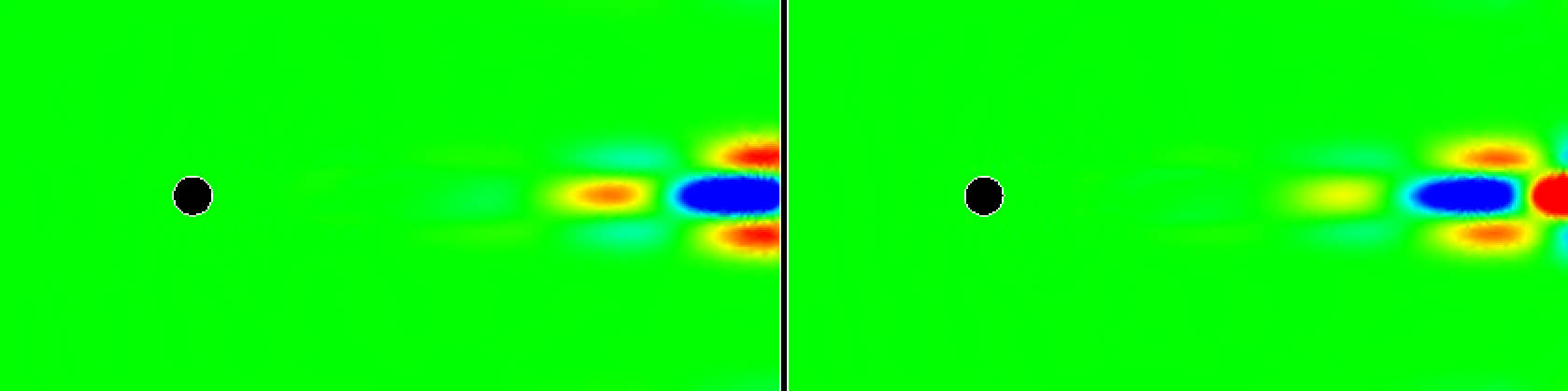}
\end{center}
\caption{Eigenmodes (real part: left, imaginary part: right) 
associated with the mean flow of the nonlinear actuation model 
displayed in Fig.\ \ref{fig:Flow_control}.
From top to bottom: the vorticity at iteration 1, 3 and 10.}
\label{fig:Flow_control_modes} 
\end{figure}
\begin{figure}
\begin{centering}
 \begin{minipage}[t]{1.2\textwidth}
\includegraphics[width=0.8\textwidth]{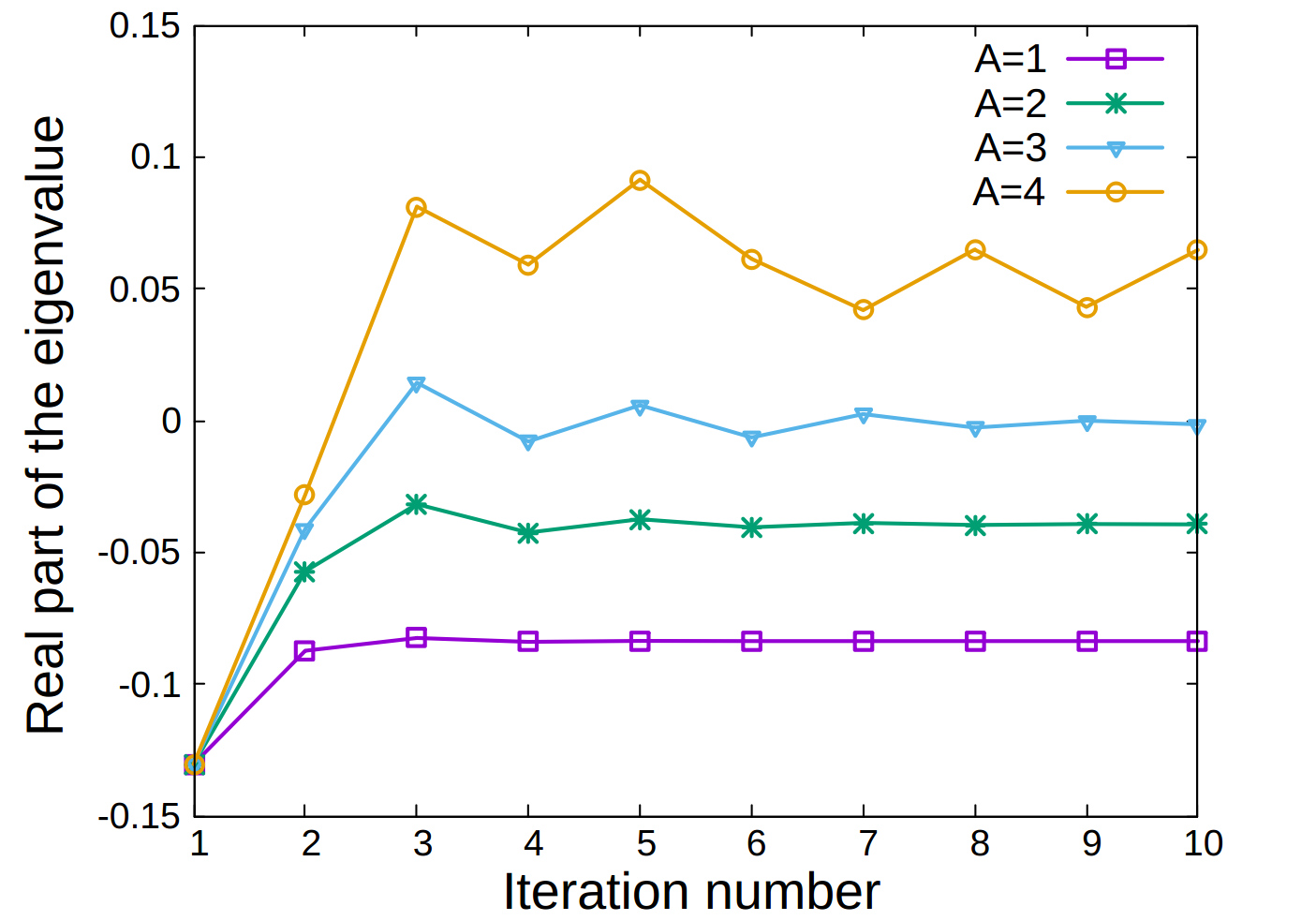}
\end{minipage}
\end{centering}
\caption{Growth rates of the eigenmodes depicted in Fig.\ \ref{fig:Flow_control_modes} in dependency of the iteration number. 
}
\label{fig:Flow_control_growthrate} 
\end{figure}

In this section,
we  demonstrate how nonlinear actuation performance can be predicted
using the unstable periodically forced Navier-Stokes solution 
computed in the previous sections. 
Starting point is the high-frequency actuation
of a circular cylinder wake at $Re=100$ of Sect.~\ref{subsec:30}. 
An experiment \cite{Thiria2006jfm} 
demonstrated that this actuation can stabilize the cylinder wake
via nonlinear frequency crosstalk mechanisms.
In the following, this nonlinear actuation effect is resolved
in a first-principle reduced-order model based on mean-field considerations.

Most  flow control experiments rely 
on the nonlinear interaction between the mean flow 
and unsteady fluctuation \cite{Tadmor1513}.  
Fluctuations change the mean flow via the Reynolds stress. 
Inversely, mean flow variations change 
also the stability properties of the fluctuations. 
This frequency crosstalk mechanism is exploited 
in a stabilization of the target instability
with low- and high-frequency actuation \cite{Brunton2015amr}.
A linearly stable solution of the Reynolds equation 
implies that the actuated flow structures 
have successfully stabilized the flow.

The stabilizing effect of  periodic actuation 
is described by the Reynolds equation
\begin{equation}
\bar{V}_{i,j} \bar{V}_{j} + \bar{P}_{,i} -\frac{1}{Re}{\bar{V}_{i,jj}} - \frac{1}{2} A^2 \nabla \cdot
  [  \vec{V}_{real} \otimes \vec{V}_{real}
 +   \vec{V}_{imag} \otimes \vec{V}_{imag} ]  = 0
\label{Eqn:Reynolds}
\end{equation}
Here $\bar{V}_i$ and $\bar P$ are the mean field values, 
$\vec{V}_{real}$ and $\vec{V}_{imag}$ are the real and imaginary parts
of the actuation mode computed with the procedure described in Sect.~\ref{subsec:31}. 
and $A$ is the  amplitude of the actuation.
It should be noted that \eqref{Eqn:Reynolds} 
only includes the Reynolds stress of the actuation mode. 
Thus, \eqref{Eqn:Reynolds} implies a successful flow stabilization
which can a posteriori be validated with the stability of computed mean flow. 

The Reynolds equation \eqref{Eqn:Reynolds} 
depends on the actuation mode $\vec{V}$
as sole contribution to the Reynolds stress.
Inversely, the actuation mode depends on the base flow 
as evidenced by the disturbance equation \eqref{Eqn:Dist}.
This reciprocal dependency between mean flow and actuation mode 
is resolved by an iteration algorithm.
In the first step, the steady Navier-Stokes solution 
is computed and taken as base flow for \eqref{Eqn:Dist} 
and subsequent eigenproblems for the actuation mode.
In the second step, 
a  base flow is updated with the Reynolds equation \eqref{Eqn:Reynolds}
using the actuation mode.
In addition, the actuation mode is re-computed with the new base flow.
These iterations are continued until convergence.
In each iteration, a global stability analysis 
may assess the suppression of the target instability.

The iterative procedure is depicted 
for forcing  $\lambda_{Im}=4.0$ encoded in the imaginary part of the eigenvalue 
and the amplitude of oscillation $A=3$.
Figs.~\ref{fig:Flow_control},
\ref{fig:Flow_control_modes}
and \ref{fig:Flow_control_growthrate}
show the convergence of the base flow, 
the actuation mode, 
the most unstable eigenmode 
and its corresponding growth-rate.

We start with the steady flow solution at $Re=100$. 
The vorticity of the flow is depicted in the first row of Fig.~\ref{fig:Flow_control} (left).
The global stability analysis of this flow field yields
von K\'arm\'an vortex shedding (Fig.~\ref{fig:Flow_control_modes}, top row) 
with negative $\lambda_{Re}$ (Fig.~\ref{fig:Flow_control_growthrate}) 
 indicating that the base flow is unstable.
The flow response to a stabilizing high-frequency rotary oscillation is resolved by  
an actuation mode,
following an earlier numerical study.
With the technique presented in Sect.~\ref{subsec:31}, 
we obtain the complex actuation mode 
shown in the top row of Fig.~\ref{fig:Flow_control} (right).

In the second iteration,
the actuation mode is employed in the Reynolds equation \eqref{Eqn:Reynolds}  
to derive the corresponding mean flow (Fig.~\ref{fig:Flow_control}, 2nd row, left).
This mean flow is used to revise the actuation mode (Fig.~\ref{fig:Flow_control}, 2nd row, right).

The process converges in about ten iterations
yielding the final mean flow field and associated actuation mode 
(Fig.~\ref{fig:Flow_control}, last row).
The stability of the base flow is assessed via a global stability analysis. 
If the growth rate is non-positive, a stabilization of the flow is achieved.

In Fig.~\ref{fig:Flow_control_growthrate}, 
the evolution of the growth rate is depicted 
at  different amplitudes of the actuation amplitude $A$,
including a strongly, marginal  and partially stabilizing level.
In Fig.~\ref{fig:Flow_control_modes},
the corresponding global modes of the mean flow 
are illustrated for subsequent iterations, 
showing the stabilization of the flow. 
The center of vorticity moves towards the outflow, 
indicating that actuation effectively  stabilizes the near-wake 
and a marginally stable von K\'arm\'an mode fluctuation may only exists far downstream.

The analysis is consistent 
with nonlinear Navier-Stokes simulations 
and serves as the demonstration of capability of the method. 
The nonlinear first-principle reduced-order model
from the iterative procedure
may resolve the performance 
of different actuation types, frequencies and amplitudes. 
Arbitrary kinds of the actuation can be tested 
using the fact that the equation  \eqref{eq:diffEV} 
and the resulting discrete one \eqref{eq:EV_Re_Im} are linear 
and we can superpose their solutions. 
At the same time we can investigate also the effect of actuator placement.
We emphasize that the resulting reduced-order model
is consistent with generalized mean-field theory 
but forced and unstable harmonic components are  fully resolved
in a first-principle Navier-Stokes discretization.
In contrast, the generalized mean-field model of \cite{Luchtenburg2009jfm}
employs empirical shedding and actuation modes, 
i.e.\ is only consistent with the one actuation for the simulation data.

\section{Conclusions}
\label{sec:6}
Several applications of frequency domain computations, 
employing complex shift have been presented. 
The method is very similar to global (fully two--dimensional or fully three--dimensional) stability analysis, 
yet yields a broader range of results. 
First,  the most amplified global eigenmodes are reproduced.
Second,  we compute unstable Navier-Stokes solutions 
featuring the periodic flow responses at any given frequency,
potentially corresponding to damped eigenmodes. 
The latter addresses a challenge of computational stability analysis:
While the most amplified eigenmode is easily distilled by a power method,
the more damped modes live 'under the radar' of many iterative solvers for eigenmodes.

The frequency domain computation has a straight-forward mathematical formulation. 
The numerical realization and the algorithm behavior demand careful 
handling of boundary conditions, meshing and numerical schemes. 
The similarity of our computations to the methods used in computational aeroacoustics 
suggests careful use of sponge-layer boundaries, 
high-order schemes and sensitivity to spurious oscillatory solutions. 
There is much less experience and techniques for these computations 
as computations in frequency domain are far more rare in 
computational fluid dynamics 
than regular time-stepping approaches.
Shifted problems are different from singular ones 
but still the numerical properties of the system are  less predictable than for the regular CFD ones.

The proposed method has evident applications in active flow control.
It could be used to explore excitable coherent structures 
which mitigate the target instability 
via nonlinear frequency crosstalk \cite{Brunton2015amr},
as it is much easier to excite than to damp an eigenmode.
The presented nonlinear first-principle reduced-order model
of wake stabilization with high-frequency forcing underlines this perspective.

The methodology can be significantly extended and has many more applications.
For example, the adjoint analysis in the spirit of Luccini \cite{Luchini2014}
is possible with minor modification of the program. 
Another application is  the use of mean flow as the base solution, 
as pioneered by Michalke \cite{Michalke1965jfm} 
and continued by  Barkley \cite{Barkley2006ep}. 
Another direction is the exploration of the rich nonlinear dynamics 
from interactions between eigenmodes  \cite{Noack2008jnet}.
These are only a few directions of future directions of research.

The presented results  are far from being a fool-proof methodology
for computation of eigenmodes 
but rather a start of a highly promising avenue
towards physical flow expansions, 
nonlinear first-principle Galerkin models 
and control applications---even for strongly nonlinear frequency crosstalk.

\begin{acknowledgement}
The authors acknowledge support by the
Polish National Science Center (NCN) under the Grant
No.: DEC-2011/01/B/ST8/07264 and by the Polish National
Center for Research and Development under the Grant No.
PBS3/B9/34/2015
and travel support of the Bernd Noack Cybernetics Foundation.
\end{acknowledgement}
\

\end{document}